\documentstyle[12pt,fleqn,epsf]{article}

\newcommand{\beq}{\begin{eqnarray}}
\newcommand{\eeq}{\end{eqnarray}}

\newcommand{\btem}{\bibitem}

\begin{document}
\begin{titlepage}
\begin{flushright}

\end{flushright}

\vspace{2cm}

\begin{center}
{\large \bf
MAXIMUM ENTROPY ANALYSIS \\
OF THE SPECTRAL FUNCTIONS IN LATTICE QCD}\\[50pt]
M. Asakawa$^{(1)}$, T. Hatsuda$^{(2)}$ and Y. Nakahara$^{(1)}$ \\  
$^{(1)}$ Department of Physics, 
Nagoya University, Nagoya 464-8602, Japan\\
$^{(2)}$ Department of Physics, University of Tokyo, Tokyo 113-0033, Japan\\


\vfill
{\bf Abstract} \\
\end{center}
\noindent
First principle calculation of the QCD spectral functions (SPFs)
based on  the lattice QCD  simulations is reviewed.
Special emphasis is placed on the Bayesian inference theory
and the Maximum Entropy Method (MEM), which is a useful tool to
extract SPFs from the imaginary-time
correlation functions numerically obtained by the Monte Carlo method. 
Three important aspects of MEM are (i) it does not require a priori
assumptions or parametrizations of SPFs, (ii) for given data, 
a unique solution 
is obtained if it exists, and (iii)
the statistical significance of the solution can be quantitatively analyzed. 
    
The ability of MEM is explicitly demonstrated by using mock
data as well as lattice QCD data.
When applied to lattice data, MEM correctly reproduces
the low-energy resonances and shows the existence of  high-energy
continuum in hadronic correlation functions.
This opens up various possibilities
for studying hadronic properties in QCD  
beyond the conventional way of analyzing the lattice data.
Future problems to be studied by MEM in lattice QCD 
are also  summarized.

\vfill

\end{titlepage}

\tableofcontents

\vspace*{1cm}

\renewcommand{\thesection}{\arabic{section}}
\renewcommand{\thesubsection}{\arabic{section}.\arabic{subsection}}
\renewcommand{\theequation}{\arabic{section}.\arabic{equation}}
\renewcommand{\thetable}{\arabic{table}}
\renewcommand{\thefigure}{\arabic{figure}}

\newpage

\setcounter{equation}{0}
\setcounter{subsection}{0}

\section{Introduction}
 
 The hadronic spectral
 functions (SPFs) in quantum chromodynamics (QCD)
 play an essential role in understanding  
 properties of hadrons as well as in probing the QCD  vacuum structure.
 For example,  the total cross section
 of the  $e^+ + e^-$ annihilation
 into hadrons can be expressed by the spectral function 
 corresponding to the correlation function of the
 QCD electromagnetic current.  
 Experimental data on the cross section actually support
 the following picture:
 asymptotically free quarks are relevant at high energies, while
 the quarks are confined inside hadronic resonances
 (such as $\rho, \omega, \phi $) at low energies.
 Another example of the application of SPF
 is the QCD spectral sum rules,
 where the moments of SPF are related to the vacuum condensates
 of quarks and gluons through the operator product expansion 
\cite{svz1,svz2}.

 The spectral functions at finite temperature ($T$) and/or baryon chemical
 potential ($\mu$) in QCD, which were originally studied in ref.\cite{hk85},
 are also recognized  as a key  concept to understand the medium      
 modification of hadrons (see, e.g., \cite{gaka,ak93,hk94,shuryak}).
 The physics motivation here
 is quite similar to the  problems in condensed matter physics
 or in nuclear physics: how elementary modes of excitations
 change their characters as $T$ and/or $\mu$ is raised.
 The enhanced   low-mass $e^+e^-$ pairs   below the $\rho$-resonance
  and the 
  suppressed high-mass $\mu^+\mu^-$ pairs at $J/\psi$
 and $\psi '$-resonance 
 observed in relativistic heavy ion collisions
 at CERN SPS \cite{ceres,charm} are typical
 examples which may  indicate  spectral changes of the $q\bar{q}$ system
  due to the effect of the surrounding environment 
  (see recent reviews, \cite{rw99,JP-review}).
 
 Since the numerical simulation of QCD on a lattice is a 
 first-principle method having  remarkable 
 successes in the study of ``static'' hadronic properties 
 (masses, decay constants, $\cdots$, etc.) \cite{lattice98,tsukuba}, 
 it is desirable to extend
 its power also to extracting hadronic SPFs.
 However, Monte Carlo simulations 
 on the lattice 
 have difficulties in accessing the ``dynamical'' quantities 
 such as spectral functions and the real-time correlation functions.
 This is because
 measurements on the lattice can only be carried out
 at a finite set of discrete points in imaginary time.
 The analytic continuation from imaginary time to 
 real time using limited and noisy lattice data
 is not well-defined and is classified as an ill-posed
 problem. This is the reason why studies to extract SPFs 
 so far had to rely on  specific ans\"{a}tze  
 for the 
 spectral shape \cite{param1,QCD-TARO}.

 In this article, we review a new way of 
 extracting the spectral functions from the lattice QCD data  
 and explore its potentiality in detail. The method we shall discuss
 is the maximum entropy method (MEM), which was recently
 applied to the lattice QCD data by the present authors \cite{nah99}.  
  
   In the maximum entropy method,
 Shannon's information entropy    \cite{shannon}
 and its generalization (which we call Shannon-Jaynes entropy 
 throughout this article) play a key role. 
 After the pioneering works on the  applications of the information entropy 
  to statistical mechanics by Jaynes \cite{jaynes57} (see also 
  \cite{jaynes}) and to 
   optical image reconstruction by Frieden \cite{frieden72},
    MEM  has been widely used in many branches
     of science \cite{wu}. Applications include
  data analysis  of quantum Monte Carlo simulations in
 condensed matter physics \cite{physrep,linden} and in nuclear physics
 \cite{koonin},  image reconstruction in crystallography and 
 astrophysics \cite{wu}, and so forth.
 In the context of QCD, MEM is a method to make statistical inference
 on the most probable SPF for given Monte Carlo data
 on the basis of the Bayesian probability theory \cite{bt}.
 
 In the maximum entropy method,
 a priori assumptions or parametrizations of the spectral functions
 need not to be made. Nevertheless, for given lattice data,
  a unique solution is obtained if it exists. 
  Furthermore,   one can carry out
 error analysis of the obtained SPF and evaluate the 
 statistical significance of its structure.
 Our basic conclusion is that
 MEM works well for the current lattice QCD data and
 can reproduce low-lying resonances as well as high-energy
 continuum structure. This opens up various possibilities
 for studying hadronic properties in QCD  
  beyond the conventional way of analyzing 
  the data.\footnote{
 We should mention here that 
 there were at least two attempts aiming at
 a goal similar to this article but 
  with different methods \cite{param2}.
  Realistic analyses of SPFs for QCD, however, have not yet 
  been carried out in those methods.}
  
  \vspace{0.2cm}
  
   This article is organized as follows. \\
  In Section 2, we shall  give a general introduction
 to the structure of the SPFs in QCD.\\
 In Section 3, the MEM procedure will be discussed in detail.
 This section is written in a self-contained way  
 so that
 the readers can easily start MEM analysis without
 referring to enormous articles that have been published in other areas.
 The uniqueness of the solution of MEM is proved explicitly
  in this Section.
  A subtle point related to the covariance matrix of the 
  lattice data is also discussed.\\
  In Section 4, we present the results of the  MEM analysis
 using the mock data. It is also discussed  
  how the resultant SPF and its error
   are affected by the quality of the mock data.\\
 In Section 5, SPFs at $T=0$ 
 in the mesonic channels are extracted by MEM
 using quenched Monte Carlo
 data obtained on $20^3 \times 24$ lattice with $\beta_{lat}=6.0$.
 Detailed analyses of the spectral structure in the 
 vector and pseudo-scalar channels are given.
    Part of the results presented in Section 4 and Section 5 
 have been  previously  reported in \cite{nah99}.\\
 In  Section 6, future problems of MEM  both from technical and physical
 point of view are summarized.  \\
 In Appendix A (B),  the statistical (axiomatic)
 derivation of the Shannon-Jaynes entropy is shown. In Appendix C,
 the singular value decomposition of matrices, which is crucial
 for numerical MEM procedure, is proved.
 
\newpage

\setcounter{equation}{0}
\setcounter{subsection}{0}

\section{Sum rules for the spectral function}

\subsection{Real and imaginary time correlations}

For simplicity, we shall focus on SPFs at 
vanishing baryon chemical potential.
We start by 
defining the following two-point correlation functions at finite $T$: 
\begin{eqnarray}
\label{ret1}
 \tilde{D}^R_{\eta \eta'}(t,{\bf x})
 & = & i \ {\rm Tr}\left(  {\rm R} \left[ J_{\eta}(t,{\bf x}) \ 
 J^{\dagger}_{\eta'}(0,{\bf 0}) \right] e^{-H/T} \right) /Z \\ \nonumber
 &  \equiv &
  \int {d^4 k \over (2 \pi)^4} \ D^R_{\eta \eta'} (k^0,{\bf k})  \ e^{-ikx} 
 \ \ , \\
\label{mat1}
 \tilde{D}_{\eta \eta'}(\tau, {\bf x})
 & = & {\rm Tr}\left( {\rm T}_{\tau} \left[
 J_{\eta}(\tau,{\bf x}) \ J^{\dagger}_{\eta'}(0,{\bf 0}) \right]
 e^{-H/T} \right) /Z \\ \nonumber
 & \equiv & T \sum_n \int {d^3 k \over (2 \pi)^3} 
\   D_{\eta \eta'} (i\omega_n,{\bf k})  \ 
 e^{-i (\omega_n \tau - {\bf k} \cdot {\bf x} ) } 
 \ \ . 
\end{eqnarray}
 Here   $J_{\eta}$ and $J_{\eta'}^{\dagger}$ 
 are composite operators in the 
 Heisenberg representation in real (imaginary) time for
 $\tilde{D}^R$ ($\tilde{D}$).
  $\eta$ and $\eta'$ denote possible Lorentz and internal indices
 of the operators.
  $Z$ is the partition function of the system defined as
 $Z={\rm Tr}  e^{-H/T} $.
 $\omega_n$ is the Matsubara frequency defined by
 $\omega_n = 2 n  \pi T$ when 
  $J_{\eta}$ and $J_{\eta'}^{\dagger}$ 
 are the Grassmann even ( = bosonic) operators and 
 $\omega_n = (2 n +1)  \pi T$ when 
 they are  Grassmann odd ( = fermionic)  operators.
  The retarded product R and the imaginary-time ordered product
  T$_{\tau}$ are defined, respectively, as
\begin{eqnarray}
 {\rm R}[A(t)B(0)] & \equiv & \theta (t)(A(t)B(0) \mp B(0)A(t)) , \\
 {\rm T}_{\tau}[A(\tau)B(0)] & \equiv &  \theta (\tau) A(\tau)B(0) \pm
 \theta (-\tau) B(0)A(\tau).  
\end{eqnarray}
  The upper sign is for bosonic  operators
  and the lower sign is for fermionic operators,
  which is used 
  throughout this article. 
Eq.({\ref{ret1}) and eq.({\ref{mat1}) are called the 
 retarded correlation and the Matsubara correlation, respectively.

 One can of course define 
 two-point correlations for more general  composite operators, but 
     $J_{\eta}$ and $J_{\eta'}^{\dagger}$ are sufficient for
 the later discussions in this article.

\subsection{Definition of the spectral function}

 The spectral function (SPF) is defined as
 the imaginary part of the Fourier transform of the retarded correlation
(see, e.g., \cite{no}),
\begin{eqnarray}
\label{def-spf} 
 & & \! \! \! \! \! \! \! \! \! \! \! \! \! \! \! \! \! \! \! \! \! \! 
 \! \! \! \! \! \! \! \! \! \! 
\hspace*{-0.6cm}
 A_{\eta \eta'} (k_0,{\bf k}) \equiv{1 \over \pi}\ 
 {\rm Im}\  D^R_{\eta \eta'} (k_0, {\bf k}) \\
   & \ \ \ \ = &  (2 \pi)^3 
 \sum_{n,m} {e^{-E_n/T} \over Z}
  \langle n | J_{\eta}(0) | m \rangle
 \langle m | J^{\dagger}_{\eta'} (0) | n \rangle (1 \mp e^{-P^0_{mn}/T})
 \delta^4 ( k^{\mu} - P^{\mu}_{mn} ) \label{finiteTspf1}\\
 & \ \ \ \ 
\stackrel{T\rightarrow 0}{\rightarrow} &  (2 \pi)^3  
 \sum_{m} \langle 0 | J_{\eta}(0) | m \rangle
 \langle m | J^{\dagger}_{\eta'} (0) | 0 \rangle 
 \delta^4 ( k^{\mu} - P^{\mu}_{m} ) ,
\label{zeroTspf1}
\end{eqnarray}
where $|n \rangle $ 
is a hadronic state with 4-momentum $P^{\mu}_n$ and the
normalization
$\langle m | n \rangle = \delta_{nm}$.
$P_{mn}^{\mu}$ is defined by $ P^{\mu}_m - P^{\mu}_n$.
$D^R (k_0, {\bf k})$ is a Fourier transform of
$\tilde{D}^R(t,{\bf x})$ with respect to both time and space
coordinates.
In eq.(\ref{zeroTspf1}), the energy of the vacuum is
renormalized to zero.

 When $\eta = \eta'$, the SPF has special properties.
 First of all, it is positive semi-definite for positive frequency:
\beq
\label{sym1}
A_{\eta \eta} (k_0 \ge 0, {\bf k}) \ge 0 .
\eeq
Also, it has a symmetry under the change of variables:
\beq
\label{sym2}
A_{\eta \eta} (-k_0, -{\bf k}) 
 = \mp A_{\eta \eta} (k_0, {\bf k}) 
 = \mp A_{\eta \eta} (k_0, -{\bf k}) ,
\eeq
where we have assumed  parity invariance of the system.

 As already mentioned in Section 1, $A_{\eta \eta'}$ is
 in some cases directly related to the 
 experimental observables. For example, consider  
  $J_{\eta} = j^{em}_{\mu}$  and
 $J_{\eta'}^{\dagger} = j^{em}_{\nu}$ with 
\beq
j^{em}_{\mu} =
 {2 \over 3} \bar{u} \gamma_{\mu} u - {1 \over 3} \bar{d} \gamma_{\mu} d
 - {1 \over 3} \bar{s} \gamma_{\mu} s \cdots ,
\eeq
which is the electromagnetic current in QCD. Then, 
  the standard $R$-ratio in the $e^+ + e^-$ annihilation 
  is related to the SPF at $T=0$ (eq.(\ref{zeroTspf1})) as follows
 (see, e.g., \cite{svz2}):
 \begin{eqnarray}
 R(s) = - { 4 \pi^2 \over s} \sum_{\mu = 0}^{3} A_{\mu}^{\mu} (s),
\end{eqnarray}
 with $s \equiv k_0^2 - {\bf k}^2$.
 The dilepton production rate from hot matter
 at finite $T$, which is an observable 
 in  relativistic heavy-ion collisions, is written through
 eq.(\ref{finiteTspf1}) as follows (see, e.g., \cite{rw99}):
 \begin{eqnarray}
 {dN_{_{l^+ l^-}} \over  d^4x d^4 k}
 = - {\alpha^2 \over 3 \pi^2 k^2}
{ \sum_{\mu=0}^{3} A_{\mu}^{\mu}(k_0,{\bf k}) 
 \over e^{k_0 /T} -1},
 \label{dileptonrate}
 \end{eqnarray}
where $\alpha$ is the electromagnetic fine structure constant. 
  The mass of leptons is neglected in this formula for simplicity.
Note that this equation is valid 
 irrespective of the state of the system, i.e., whether
it is in the hadronic phase or in the quark-gluon plasma.

\subsection{Dispersion relations}

Using  the definition of the two point 
 correlations eqs.(\ref{ret1},\ref{mat1}) together with the
 general form of 
 the spectral function eq.(\ref{def-spf}), one can derive the 
 dispersion relation in the momentum space  \cite{no};
\beq
\label{ret2}
D^R_{\eta \eta'}(k_0, {\bf k}) 
= \int_{-\infty}^{+\infty}
  d\omega \ {A_{\eta \eta'}(\omega,{\bf k})
  \over \omega - k_0 - i \epsilon} - ({\rm subtraction}), \\
\label{mat2}
D_{\eta \eta'}(i \omega_n, {\bf k}) 
=   \int_{-\infty}^{+\infty}
 d\omega \ {A_{\eta \eta'}(\omega,{\bf k})
  \over \omega - i \omega_n} - ({\rm subtraction}).
\eeq
 The integrals on the right hand side (r.h.s.) of eqs.(\ref{ret2},\ref{mat2})
  do not always converge and need appropriate 
 subtractions. This is because, in QCD,  
\beq
\label{a-asympt}
A_{\eta \eta'}(\omega \rightarrow \infty, {\bf k})
   \propto \omega^{n},
\eeq
  where $n = {\rm dim} [J_{\eta}] 
 + {\rm dim}[J_{\eta'}] -4$, with 
 ${\rm dim}[J]$ being  the canonical dimension of the
 operator $J$. For example, $n=2$ for the mesonic correlation
 with $J \sim \bar{q} q$ and $n=5$ for baryonic correlation 
 with $J \sim qqq$.

\subsection{Sum rules}

 From the dispersion relations eqs.(\ref{ret2},\ref{mat2}),
 one can derive several   {\em sum rules} for the spectral functions.
 
For example, for bosonic operators with  $\eta'=\eta$,  
 eq.(\ref{ret2}) is rewritten as
 \beq 
\label{qcdsr1}
 D^R_{\eta \eta}(k^0, {\bf k}) 
= \int_{0}^{\infty} d\omega^2 \  
{A_{\eta \eta}(\omega,{\bf k})
  \over \omega^2 - k_0^2 - i \epsilon} - ({\rm subtraction}).
\eeq
In  the deep-Euclidean region  $k_0^2 \rightarrow -\infty$
with finite  ${\bf k}$, we make
 the operator product expansion of the left hand side (l.h.s.) of
 (\ref{qcdsr1}) and
subsequently apply  the following Borel transformation on both sides of
 eq.(\ref{qcdsr1}):
\begin{equation}
B_M \equiv
\lim_{\scriptsize
\begin{array}{l}
-k_0^2,n\rightarrow \infty\\
-k_0^2/n=M^2
	\end{array}}
\left (-k_0^2\right ) ^n
\left ( \frac{d~}{dk_0^2} \right ) ^n \ \ \ .
\end{equation}
 Then one arrives at the Borel sum rule in the medium
 \beq
 \label{medium-QSR}
\! \! \! \! \! \! \! \! \! \! \! \! \! \! 
\int_0^{\infty} 
A_{\eta \eta}(\omega,{\bf k}) e^{-\omega^2/M^2} d\omega^2 
= M^{2{\rm dim}{J}- 2} \sum_{n,m=0}^{N}
 C_{nm}(\alpha_s(M^2))  
   {\langle\!\langle {\cal O}_{2n}(M^2) \rangle\!\rangle
    \over M^{2n}}  
        \left( {{\bf k}^2 \over M^2} \right) ^m .
\eeq
Here the r.h.s. of  (\ref{medium-QSR}) is an
asymptotic expansion in terms of $\Lambda_{QCD}^2/M^2$,
$T^2/M^2$ and ${\bf k}^2/M^2$ with 
$M^2$ being the Borel mass. Therefore, $M^2$ should be 
large enough for the expansion to make sense.
$C_{nm}$ are the dimensionless coefficients 
calculated perturbatively with the QCD
running coupling constant $\alpha_s(M^2)$
defined at the scale $M^2$. 
$\langle\!\langle {\cal O}_{2n}(M^2) \rangle\!\rangle $
is the thermal average of all possible 
local operators with canonical dimension $2n$
renormalized at the scale  $M^2$.
Note that $T$ enters  only through
the matrix elements $\langle\!\langle {\cal O}_{2n}(M^2) \rangle\!\rangle $.
 ${\cal O}_{2n}$ contains not only the Lorentz scalar operators
but also Lorentz tensor operators because of 
 the existence of a preferred frame 
 at finite $T$.
In an appropriate Borel window  
($M_{min} < M < M_{max}$), (\ref{medium-QSR}) gives 
constraints on the integrated spectral function.
The in-medium sum rule in the form, eq.(\ref{medium-QSR}), 
was first derived in \cite{HL}. It is a natural generalization
of the original QCD sum rule in the vacuum \cite{svz1,svz2}.
Recent applications of the in-medium QCD sum rules can be found in
\cite{Lee-rev}.  

Another form of sum rule, which is closely
related to
 the main topic of this article,  is derived from eq.(\ref{mat2}). 
 Let us define the mixed representation of the Matsubara correlator,
 \beq
 D_{\eta \eta'}(  \tau  , {\bf k} ) 
 & = &  T \sum_n e^{-i \omega_n \tau} D_{\eta \eta'}(i\omega_n, {\bf k} )  \\
 \nonumber
 & = &  \int_{-\infty}^{+\infty} d\omega \  
  A_{\eta \eta'}(\omega,{\bf k}) 
 \left[ 
   T \sum_n {e^{-i\omega_n \tau} \over \omega  - i \omega_n} \right] .
\eeq
 Then, by using the identity,
\beq
 T \sum_n  {e^{-i\omega_n \tau} \over x - i \omega_n}
 = {e^{-x \tau} \over 1 \mp e^{-\beta x}} \ \ \ (0 \le   \tau < \beta ),
\eeq
one arrives at the sum rule
\beq
\label{KA}
 D_{\eta \eta'}( \tau , {\bf k} ) 
 = \int_{-\infty}^{+\infty} {e^{-\tau \omega} \over
 1 \mp e^{-\beta \omega}} \  A_{\eta \eta'} (\omega, {\bf k})
 \ d\omega  \ \ \ \ \ \ \ \ \ (0 \le   \tau < \beta ).
\eeq

 Eq.(\ref{KA}) is always convergent and does not require
 subtraction as long as
 $\tau \neq 0$. This is because
 $A(\omega, {\bf k})$ in QCD has at most power-like
 behavior at large $\omega$ (see eq.(\ref{a-asympt})).
 Owing to this, we always
 exclude the point $\tau=0$ in the following analysis.

 When ${\eta }={\eta'}$  and $J_{\eta}$ is a quark bilinear operator 
 such as $\bar{q} q$, $\bar{q} \gamma_{\mu} q$, $\cdots $
  etc.,
  one can further reduce the sum rule 
 into a form similar to the Laplace transform:
 \beq
 \label{KAB}
 D( \tau ,{\bf k}) 
& = & 
\int \tilde{D}( \tau , {\bf x})\ e^{-i{\bf k}\cdot {\bf x}}\  d^3x
  \nonumber \\ 
 & = &  \int_{0}^{+\infty} { e^{-\tau \omega} + e^{-(\beta - \tau) \omega }
  \over
 1 - e^{-\beta \omega} } \  A (\omega,{\bf k}) \ d\omega \nonumber \\
 & \equiv &  
  \int_{0}^{+\infty} \ K(\tau,\omega)  \  A (\omega,{\bf k}) \ d\omega 
 \ \ \ (0 \le   \tau < \beta ), 
\eeq
where we have suppressed the  index  $\eta$ for simplicity.
$K$ is  the kernel of the integral transform.
Eq.(\ref{KAB}) 
is the basic formula, which we shall utilize later.
From now on, we focus on SPFs ``at rest'' (${\bf k}=0$) 
for simplicity and
omit the label ${\bf k}$. 

 To get a rough idea about the structure of 
  $D(\tau)$, let us consider a parametrized form of  SPF, $A_{_V}(\omega)$,
 for the correlation of isospin 1  vector currents. It consists of
  a pole (such as the $\rho$-meson) + continuum:
\beq
\label{p+c}
A_{_V} (\omega) &  = &  \omega^2 \rho_{_V} (\omega),  \nonumber \\
\rho_{_V} (\omega) & = &
 2F_{_V}^2 \ \delta (\omega^2 - m_{_V}^2) + {1 \over 4 \pi^2}
 \left (1 + {\alpha_s \over \pi} \right ) \theta (\omega - \omega_0) .
\eeq
 Here $m_{_V}$ and $\omega_0$ are the mass of the vector meson and
 the continuum threshold, respectively. $F_{_V}^2$ is the residue at the vector
 meson pole. This simple parametrization of  SPF has been commonly used in 
 the QCD sum rules \cite{svz1,svz2}. (For a more realistic 
 parametrization, see eq.(\ref{realistic}) in Section 4.2.)
 $D(\tau)$ at $T=0$ corresponding to (\ref{p+c}) reads
\beq
\label{sh-d}
D(\tau) = F_{_V}^2 m_{_V} e^{-m_{_V} \tau} 
 + \left (1 + {\alpha_s \over \pi} \right ) 
 \left( {2 \over \tau^3} + {2 \omega_0 \over \tau^2}
         + {\omega_0^2 \over \tau} \right) e^{-\omega_0 \tau}.
\eeq
At long distances (large  
 $\tau$), the exponential decrease of (\ref{sh-d}) 
 is dominated by the pole contribution. On the other hand,
 at short distance (small $\tau$),  the power behavior $1/\tau^3$
 from the continuum dominates eq.(\ref{sh-d}).
 Although (\ref{sh-d}) captures some essential parts,
 $D(\tau)$ in the real world or on the lattice has richer
 structure. Studying this without any ans\"{a}tze like
 (\ref{p+c}) is a whole aim of this article.

\subsection{An ill-posed problem}

Monte Carlo simulation provides  $D(\tau )$ in (\ref{KAB}) 
for a discrete set of points, $\tau = \tau_i$, with
\begin{eqnarray}
1 \le \tau_i/a   \le N_\tau ,
\end{eqnarray}
where $N_\tau$ is the number of the temporal lattice sites and $a$ is the 
lattice spacing.
 In the actual analysis, we use data points in a limited domain
 $\tau \in [ {\tau}_{min}, {\tau}_{max} ]$.
 Since we can generate only finite number of gauge
 configurations numerically, the lattice data $D(\tau_i)$ 
 has  a statistical error.
 From such finite  number of  data  with noise, 
 we need to reconstruct the 
 continuous function $A(\omega)$ on the r.h.s. of
 eq.(\ref{KAB}), or equivalently to perform the inverse Laplace transform.

This is a typical ill-posed problem, where
the number of data points  
  is much smaller than the number of
degrees of freedom to be reconstructed.
 The standard 
 likelihood analysis ($\chi^2$-fitting)  is obviously inapplicable here, 
 since  many degenerate solutions  appear in the process of  minimizing
  $\chi^2$.
 This is the reason why the previous analyses of the spectral functions   
 have been done only under strong assumptions on the spectral shape 
  \cite{param1,QCD-TARO}.
  Drawbacks of the previous approaches are twofold: 
 (i) a priori assumptions on SPF prevent us from studying the
 fine structures of SPF, and (ii) the result does not have
 good stability against the change of the number of parameters
 used to characterize SPF. Both disadvantages become even more
 serious at finite $T$, where we have almost
 no prior knowledge on the spectral shape.

The maximum entropy method, which we shall discuss in the next section,
is a method to circumvent these difficulties
by making a statistical inference of the most probable
 SPF (or sometimes called  the {\em image} in the following)
 as well as its reliability on the basis of a 
 limited number of noisy data.

\newpage

\setcounter{equation}{0}
\setcounter{subsection}{0}

\section{Maximum entropy method (MEM)}

 In this section, we shall discuss the MEM procedure
 in some detail to show its basic principle as well as to show  
 crucial points in its application to lattice QCD data.

 The theoretical basis of the maximum entropy method
 is  Bayes' theorem in probability theory \cite{bt}:
\begin{eqnarray}
\label{bayes_t}
P[X|Y] = {P[Y|X]P[X] \over P[Y]} \ \ ,
\end{eqnarray}
 where $P[X|Y]$ is the conditional probability of $X$ given $Y$.
 The theorem is easily proved by using the product formula
 $P[XY] = P[X|Y] P[Y] = P[Y|X]P[X]$. 
 Let $D$ stand for Monte Carlo data with errors for a specific channel
 on the lattice and $H$ summarize all the definitions and
 prior knowledge such as  $A(\omega \ge 0) \ge 0$.
 From  Bayes' theorem,
 the conditional probability of having 
 $A(\omega)$ given the data reads 
\begin{equation}
\label{bayes_latt}
P[A|DH] = {P[D|AH]P[A|H] \over P[D|H]} .
\end{equation}
Here $P[D|AH]$ and $P[A|H]$ are called the {\em likelihood function}
 and the {\em prior probability}, respectively.
 $ P[D|H]$ is simply a normalization constant independent of $A(\omega)$.
 (Note here that it may be more appropriate to call  $P$
 ``plausibility'' instead of   ``probability'', since
 $P$ does not necessarily
 have the frequency interpretation \cite{jaynes,smith}.)

Now, the most probable image is $A(\omega)$ that satisfies
the condition,
\beq
\label{stationary}
{ \delta P[A|DH] \over \delta A} =0.
\eeq
Furthermore, the reliability of the image satisfying 
(\ref{stationary}) can be estimated by the 
second variation, or schematically,
${ \delta^2 P[A|DH] / \delta A \delta A}$.

 When more data become available, 
 $P[A|DH]$ can be updated.
 This is seen from the following chain rule, which is
 a consequence of  Bayes' theorem and the  product formula
 for conditional probabilities:
\begin{eqnarray}
P[A|D_1 D_2 H] = P[D_2|D_1 A H] P[D_1 |A H] P[A|H] /P[D_1 D_2 | H].
\end{eqnarray}
For further discussions on the general use of Bayesian analysis,
see ref. \cite{bt}.

To go further, we need to specify the explicit forms
of the likelihood function and the 
prior probability. This will be discussed below.

\subsection{Likelihood function}

For large number of Monte Carlo measurements of a correlation
function, the data is expected to obey the  Gaussian distribution
according to the central limit theorem:
\begin{eqnarray}
\label{chi2}
P[D|AH] & = & {1 \over Z_L}\ e^{-L}\ \ , \\
\label{chi2-1}
L & =& {1 \over 2} \sum_{i,j}
(D(\tau_i)-D_A(\tau_i))C^{-1}_{ij} (D(\tau_j)-D_A(\tau_j)),
\end{eqnarray}
where $i$ and $j$ run over the
actual data points which we utilize in the analysis,
$ {\tau}_{min}/a \leq i,j \leq {\tau}_{max}/a$.  
For later purposes, we define the number of   data points to be used in MEM,  
\begin{eqnarray}
N={\tau}_{max}/a - {\tau}_{min}/a+1.
\end{eqnarray}
$D(\tau_i )$ is the lattice data averaged over gauge configurations, 
\begin{eqnarray}
\label{average-D}
D(\tau_i)= {1 \over N_{conf}} 
\sum_{m=1}^{N_{conf}} D^{m}(\tau_i) ,
\end{eqnarray}
where $N_{conf}$ is the total number of gauge configurations
and $D^{m}(\tau_i)$ is the data for the $m$-th gauge configuration.
$D_A(\tau_i )$ in (\ref{chi2-1}) is the correlation function defined 
by the r.h.s. of eq.(\ref{KAB}).

$C$ in (\ref{chi2-1}) is an $N \times N$ covariance matrix defined by
\begin{eqnarray}
\label{covariance}
C_{ij}= {1 \over N_{conf}(N_{conf}-1)} 
\sum_{m=1}^{N_{conf}}
(D^{m}(\tau_i)-D(\tau_i))(D^{m}(\tau_j)-D(\tau_j)).
\end{eqnarray}
Lattice data have generally strong correlations among
different $\tau$'s, and it is essential to take into account the
off-diagonal components of $C$.

The integration of $P[D|AH]$ over $D$ with the measure $[dD]$ defined below
is normalized to be unity.
$Z_L$ denotes the corresponding normalization constant:
\begin{eqnarray}
\label{L-normalization}
[dD ] \equiv \prod_{i=\tau_{min}/a}^{\tau_{max}/a} dD(\tau_i), \ \ \ 
Z_L = (2\pi)^{N/2} \sqrt{\det C}.
\end{eqnarray}

 In the case where $P[A|H]=$ constant
 in eq.(\ref{bayes_latt}),
 maximizing $P[A|DH]$ is equivalent to
 maximizing eq.(\ref{chi2}) with respect to $A$, which
 is nothing but the standard $\chi^2$-fitting.
 On the lattice, as we shall see later, the number of lattice data is 
 ${\cal O}(10)$, which is 
 much smaller than the number of points of the spectral function  
 to be reproduced (${\cal O}(10^3)$). Therefore,
 the $\chi^2$-fitting does not work. This difficulty is overcome
 in the maximum entropy method, where the existence of a non-constant
 $P[A|H]$ plays an  essential role.

\subsection{Prior probability}

 In MEM, 
 the prior probability is written
with auxiliary
parameters $\alpha$ and $m$ as
\begin{eqnarray}
\label{prior}
P[A|H\alpha m]= {1 \over Z_S} e^{\alpha S},
\end{eqnarray}
where $S$ is the Shannon-Jaynes entropy,
\begin{eqnarray}
\label{entropy}
S & = & \int_0^{\infty} \left[ A(\omega ) - m(\omega ) -
A(\omega)\log \left ( \frac{A(\omega)}{m(\omega )} \right ) \right] d\omega \\
\label{entropy-dis}
& & \ \ \longrightarrow \ \ \ \sum_{l=1}^{N_{\omega}} 
\left[ A_l - m_l - A_l \log \left ( \frac{A_l}{m_l} \right ) \right] .
\end{eqnarray}
$\alpha$ is a real and positive parameter, while 
$m(\omega )$ is a real and positive function called 
the {\em default model}  or the {\em prior estimate}.
Although $\alpha$ and $m$
are a part of the hypothesis $H$ in (\ref{bayes_latt}), we
write them explicitly on the l.h.s. of 
(\ref{prior}) to separate them from the other hypotheses. 
(To be consistent with this notation, we  
replace $P[D|AH]$ in (\ref{chi2})  by   $P[D|AH\alpha m]$ in the
following, although
$P[D|AH\alpha m]$ does not depend on  $\alpha$ or $m$.) 
$\alpha$ will be integrated out later and is eliminated  in the 
final results. $m$ remains in the final results, but one can study
the sensitivity of the results against the change of $m$.
   
In the numerical analysis,
the frequency is discretized into  $N_{\omega}$ pixels of an equal size
$\Delta \omega$   as shown in (\ref{entropy-dis});  
namely, 
 $A_l \equiv A(\omega_l) \Delta \omega$ and
$m_l \equiv m(\omega_l) \Delta \omega$  with 
$\omega_l \equiv l \cdot \Delta \omega$ ($ 1 \leq l \leq N_{\omega}$). 

The integration of $P[A|H\alpha m]$ over $A$ with the
measure $[dA]$ defined below is normalized to be unity.
$Z_S$ is the corresponding normalization factor:
\begin{eqnarray}
\label{S-normalization} 
 [dA] \equiv 
  \prod_{l=1}^{N_{\omega}} {dA_l \over \sqrt{A_l}}, \ \ \
 Z_S \simeq  
  \left( {2 \pi \over \alpha } \right) ^{N_{\omega}/2} .
\end{eqnarray}

In Appendix A, we give a derivation of the Shannon-Jaynes entropy $S$,
the measure $[dA]$ and  the normalization $Z_S$, on the basis of  
 the so-called  ``monkey argument'' \cite{skill1,gd78,jaynes86}.
The prior probability with eq.(\ref{entropy}) is shown to be the
most unbiased one for  positive images.
The structure of $S$ in (\ref{entropy})  can be alternatively 
derived  on an axiomatic basis (see \cite{skill2} and references therein):
For completeness, 
we present, in Appendix B, the axioms somewhat simplified from those
   given in   \cite{skill2}.

 \subsection{Outline of the MEM procedure}
 
 The procedure of the maximum entropy method
 may be classified into three classes:
 historic, classical \cite{skill1} and Bryan's method \cite{bryan}.
 They are different in the treatment of $\alpha$ as well as 
 the way to search the maximum of $P[A|DH\alpha m]$.
 In this article, we shall follow 
 Bryan's method, which is the state-of-art MEM
 with the most efficient algorithm and 
 the least conceptual difficulty. The method consists of the steps
 given below.
 
\vspace{0.3cm}

 \noindent
{\bf Step 1:} Searching for the most probable image for a given $\alpha$.

 Combining (\ref{bayes_latt}), (\ref{chi2}) and (\ref{prior}),
 one obtains
 \begin{eqnarray}
 P[A|DH\alpha m] \propto {1 \over Z_S Z_L} e^{Q(A)} \ \ , \ \ \ \ \ 
 Q (A) \equiv \alpha S - L.
 \end{eqnarray}
 Therefore, the most probable image for a given $\alpha$ (and $m$),
 which we call $A_{\alpha}$,  
 satisfies
 \begin{eqnarray}
\label{max1}
 \left .
{\delta Q \over \delta A(\omega)}
 \right | _{A = A_{\alpha}} =0.
 \end{eqnarray}
 At this stage, $\alpha$ plays a role of a parameter which
 controls the relative weight of the
entropy  $S$ (which tends to fit $A$ to the default model $m$)
and the likelihood function $L$ (which tends to fit $A$ to the
lattice data).  

 As will be proved in Section 3.5, the solution of eq.(\ref{max1})
 is unique if it exists.
 This makes the MEM analysis robust and  essentially different from the
 $\chi^2$-fitting. The latter can have  many degenerate solutions
 in  ill-posed problems.
 Further details on the algorithm for solving (\ref{max1}) are given in 
 Section 3.4.

\vspace{0.3cm}

\noindent
{\bf Step 2:} Averaging over $\alpha$.

The final output image $A_{out}$ is
defined  by a weighted average over $A$ and  $\alpha$:
\begin{eqnarray}
\label{final}
 A_{out}(\omega)   & = &   
    \int [dA] \ \int d\alpha\ 
 \ A(\omega)  \ P[A|DH\alpha m]P[\alpha|DHm] 
 \nonumber \\
    &\simeq & \int d\alpha \ A_{\alpha}(\omega)  \ P[\alpha|DHm] \ .
\end{eqnarray}
Here, we have assumed 
that $P[A|DH\alpha m] $ is
sharply peaked around $A_{\alpha}(\omega)$, which should be
 satisfied for good data, i.e., data with small errors.  Under this
  assumption  $P[\alpha | DHm]$ can be evaluated using Bayes' theorem as
 \begin{eqnarray}
\label{prior-alpha}
 P[\alpha|DHm] & = & 
  \int [dA] P[D|AH\alpha m]P[A|H\alpha m]P[\alpha|H m]/P[D|H m] \nonumber \\
          & \propto & P[\alpha|Hm] \int [dA] \ {1 \over Z_S Z_L} e^{Q(A)}  \\
\label{prior-alpha2}
          & \propto & 
          P[\alpha|Hm] \  \exp \left[ 
            {1 \over 2} \sum_k \log {\alpha \over \alpha + \lambda_k}
           + \alpha S(A_{\alpha}) - L(A_{\alpha}) \right].
 \end{eqnarray}
Here, $\lambda_k$'s are the eigenvalues of the real symmetric matrix
in the frequency space,
\beq
  \Lambda_{l,l'} \equiv
  \left.  \sqrt{A_l}
{ \partial^2 L \over \partial A_l \partial A_{l'} }
 \sqrt{A_{l'}}  \right|_{A = A_{\alpha}}.
\eeq
The standard choice of the prior probability for $\alpha$ is  
either Laplace's rule ($P[\alpha|H m] = {\rm const.}$) or  Jeffreys' rule
($P[\alpha|Hm] = 1/\alpha$) \cite{bt}. However, 
the integral on the r.h.s. of
eq.(\ref{final}) is insensitive to 
the choice, as long as the probability is concentrated
around its maximum at $\alpha = \hat{\alpha}$. We have checked that
this is indeed the case for our lattice QCD data.
Therefore,  we use the Laplace rule  for simplicity throughout this article.

 As for the averaging over $\alpha$,
  we first determine a region of $\alpha$, 
$[\alpha_{min}, \alpha_{max}]$, by the criterion
$P[\alpha|DHm] \geq 10^{-1} \times P[\hat{\alpha}|DHm]$. 
Then, after renormalizing  $P[\alpha|DHm]$ so that
$\int_{\alpha_{min}}^{\alpha_{max}}
d\alpha \ P[\alpha|DHm] =1$ is satisfied,
we carry out the integration eq.(\ref{final}) over the above
 interval. 

In the analysis in Section 4 and Section 5, we make 
 a further approximation on $P[\alpha | DHm]$:
 {\em After} obtaining $\hat{\alpha}$
 by maximizing  eq. (\ref{prior-alpha2}), 
 $A_{\alpha}$ and $\lambda_{k}(\alpha)$  in (\ref{prior-alpha2})
 are assumed to have weak $\alpha$-dependence
 and are replaced by $A_{\hat{\alpha}}$ and $\lambda_{k}(\hat{\alpha})$
 respectively.  
 The approximate form  $P^{app}[\alpha | DHm]$ is used for carrying out
 the averaging in (\ref{final}). We have checked that this approximation does 
 not lead to an error larger
  than the line-width of the figures for SPF, although
 $P^{app}[\alpha | DHm]$ itself for large $\alpha$ differs from 
 $P[\alpha | DHm]$ by as much as 10\%.

\vspace{0.3cm}

\noindent
{\bf Step 3:} Error analysis.

 The advantage of  MEM  is that
 it enables one 
 to study the statistical significance
 of the reconstructed image $A(\omega)$ quantitatively. 
 The error should be 
 calculated for $A(\omega)$ averaged over some interval in $\omega$,
 since there are correlations among $A(\omega)$ at
 neighboring $\omega$'s.
 This is explicitly seen from the Hesse matrix of $Q$ defined by
\begin{equation}
Y_{\omega \omega'} 
= \left . \frac{\delta^2 Q}{\delta A (\omega) \delta A(\omega ')}
 \right | _{A = A_\alpha},
\end{equation}
which is not diagonal in general. 

In the following MEM analysis, the error estimate is carried out
for (weighted) averages of the image $A_{out}(\omega)$ in finite
regions instead of estimating errors at each pixel \cite{physrep}.
 For this purpose, 
 we first define the average of $A(\omega)$ given  $\alpha$,
\begin{eqnarray}
\langle A_{\alpha} \rangle_{I,W}  \equiv   
\frac{\int [dA] \ \int_{I} d\omega \  A(\omega) P[A|DH\alpha m]
W(\omega) }{\int_{I} d\omega \ W(\omega) } 
 \simeq  
\frac{\int_{I} d\omega \ A_\alpha (\omega) 
 W(\omega)  }
{\int_{I} d\omega \  W(\omega)}, \label{average}
\end{eqnarray}
where $I = [\omega_1, \omega_2]$ is a given region 
in the $\omega$-space and $W(\omega)$ is a weight function.
To get the last expression in
  eq.(\ref{average}), it is assumed as before that
 $P[A|DH\alpha m]$ is highly peaked around $A_{\alpha}(\omega)$.

For simplicity we use the flat weight, $W(\omega)=1$, and
omit the suffix $W$ from now on.
 The variance of $\langle A_{\alpha} \rangle_{I}$  is similarly 
 estimated as
\beq
\label{error1}
  \langle (\delta A_{\alpha})^2 \rangle_I & = & 
 \int [dA] \ \int_{I\times I} d\omega d \omega' \  
 \delta A(\omega) \delta A(\omega ')
 P[A|DH\alpha m] \ / \ \int_{I\times I} d\omega d \omega' \\
 & \simeq  & 
 - \int_{I\times I} d\omega d\omega' 
 \ \left(  {\delta^2 Q \over \delta A(\omega) \delta A(\omega')}
  \right)^{-1}_{A=A_{\alpha}} \ /\ \int_{I\times I} d\omega d \omega'  ,
\eeq
where $\delta A(\omega) = A(\omega) - A_\alpha (\omega)$. 
 The Gaussian approximation of 
$P[A|DH\alpha m]$ around $A_{\alpha}$ is taken in the last expression.

As we have done for $A_{out}(\omega)$,
the error for $A_{out}$ in the region $I$
is also given by the following average;
\begin{equation}
\langle (\delta A_{out})^2 \rangle_I
\equiv
\int d\alpha \ \langle (\delta A_\alpha)^2 \rangle_I
\ P[\alpha |DHm] \ \ .
\end{equation}

 By using the same procedure with a slight modification,
 it is possible to estimate the errors for
 quantities derived from the reconstructed image. One example
  is the 
 pole residue, $ \int_{pole} A_{out}(\omega)\ d\omega^2$, with
  $``pole"$ implying the integral over the 
   lowest isolated pole.
\vspace{0.3cm}

\noindent
{\bf Step 4:} Sensitivity test under the variation of $m$.

 One can study the sensitivity
 of the final results under the variation of $m$.
 In QCD, $m(\omega \rightarrow {\rm large})$ can be
 estimated with perturbation theory
 (see eq.(\ref{asympt-form})). Therefore, the value
 of $m$ is known at least at large $\omega$.
    
 When the final result is not  stable enough against the 
 variation of $m$,
 one may select the optimal $m$ that  gives the image with the smallest
 error for the spectral functions.   
  
\subsection{Maximum search using SVD}\label{search}

The non-trivial part of the MEM analysis is to find the global maximum
of $Q$ in the functional space of $A(\omega)$, which has typically
$O(10^3)$ degrees of freedom in our case.  Fortunately,
 the singular value decomposition (SVD) of the kernel $K$ reduces 
 the  search direction into a subspace with a
 dimension not more than the number of data points $\sim O(10)$.
 This  has been shown by Bryan \cite{bryan}. 
 In the following, we shall discuss the essential points
 of the algorithm and its usage in our problem.
 
Let us first recapitulate the definition of  the 
 discretization in  the frequency $\omega$ and the imaginary time
$\tau$;
\beq
 \omega_l & = &  l \cdot  \Delta \omega , \ \
 A_l = A(\omega_l)\cdot\Delta \omega ,
 \ \ \ \ \ \ \ \ \ \ (l= 1, 2, 3,
  \cdot \cdot \cdot , N_{\omega}), \\
 \tau_i   & = &  i \cdot  \Delta \tau  
 = i \cdot a , \ \ 
 D_i = D(\tau_i) ,
   \ \ \ \ \ \ \ \  (i = 1, 2, 3, \cdot \cdot \cdot , N),
\eeq
 where $\Delta \omega$ is the 
 pixel size of $O(10$ MeV) in the frequency space (see Section 5.4 on
 how to choose $\Delta \omega$) and $N_{\omega}$ is the mesh size of $O(10^3)$.
 $\Delta \tau = a$ is the lattice spacing in the temporal direction.
 We have defined $N = \tau_{max}/a - \tau_{min}/a +1 \sim {\cal O}(10)$
 as before.

 The extremum condition 
  $\partial Q / \partial A_l=0$ (eq.(\ref{max1})) 
  with eq.(\ref{KAB})  is written as
\begin{eqnarray} 
\label{br1}
 -\alpha \log \frac{A_l}{m_l} 
 = \sum_{i=\tau_{min}/a}^{\tau_{max}/a} 
 K_{il} \frac{\partial L}{\partial D_{A i}}, \ \ \ {\rm where}~~
 K_{il} = K(\tau_i, \omega_l).
 \end{eqnarray}

Since $A_l$ is positive semi-definite, it is 
parametrized as 
\begin{eqnarray}
\label{br2}
A_l = m_{_l} \exp a_{_l} \ \ \ \ (1 \leq l \leq N_{\omega}).
\end{eqnarray}
 Here $\vec{a}= (a_1, a_2, \cdot \cdot \cdot , a_{_{N_{\omega}} })^t$
 is a general column vector in the $N_{\omega}=O(10^3)$ dimensional space. 
 However, 
 $\vec{a}$ as a solution of (\ref{br1}) turns out to be confined in the 
 so called ``singular subspace'', 
whose dimension is $N_s ( \le N \ll N_\omega$).
To show this,  let us  substitute (\ref{br2}) into (\ref{br1}),
\begin{eqnarray}
\label{br3}
- \alpha \vec{a} = K^t  \overrightarrow{  \frac{\partial
L}{\partial D_A} },
\end{eqnarray}
where $K^t$ is an $N_{\omega} \times N$ matrix and
$\overrightarrow{ \frac{\partial L}{\partial D_A}}$ is
an $N$ dimensional column vector.

 The SVD of $K^t$ is defined by
 $K^t = U \Xi V^t$, where 
 $U$ is an $N_{\omega} \times N$ matrix satisfying
 $U^t U = 1$,  $V$ is an $N \times N$ matrix satisfying
 $V^t V = V V^t = 1$, and  $\Xi$ is an $N \times N$ diagonal
 matrix with positive semi-definite elements,
 $\xi_i \,( i=1,2,...,N )$, which are called the singular values
 of $K^t$  \cite{chaterin}. $\xi_i$'s may be  ordered in such a way that
 $\xi_1 \geq \xi_2 \geq \cdots \geq \xi_{N_s} > 
 \xi_{N_{s+1}}= \cdots =0$, where
\begin{eqnarray}
 N_s \equiv {\rm rank}[ K^t] \leq N \leq N_{\tau}.
\end{eqnarray}
 The proof of the singular value decomposition
  is given in Appendix C for completeness.
 For the kernel such as 
 $(K^t)_{li} = {\rm exp}(-\omega_l \tau_i)$,
 the singular values $\xi_j$ are all non-zero but 
 decrease exponentially as $j$ increases.
 
The explicit form of SVD is written as
\begin{eqnarray}
\label{br4}
K^t &=& U \, \Xi \, V^t \nonumber \\
&=& \left( 
\begin{array}{ccc}
u_{_{11}} & \ldots & u_{_{1 N}} \\
&&\\
\vdots & \ddots & \vdots \\
&&\\
u_{_{N_\omega 1}} & \ldots & u_{_{N_\omega N}} 
\end{array}
\right)
\left(
\begin{array}{lcr}
\xi_{_1}\,0& \ldots & 0~~\\
0 & \null & \null~~\\
\vdots & \ddots\ddots\ddots & \vdots~~\\
\null& \null & 0~~\\
0 & \ldots & 0\,\xi_{_N}\\
\end{array}
\right)
\left(
\begin{array}{ccc}
v_{_{11}} & \ldots & v_{_{N1}}\\
\vdots & \ddots & \vdots \\
v_{_{1N}} & \ldots & v_{_{NN}}
\end{array}
\right) \ .
\end{eqnarray}
Following Bryan \cite{bryan},  
we define the $N_s$ dimensional space spanned by
the first $N_s$ columns of $U$ as the ``singular space''.
The bases in this space
are $\{ \vec{u}_1, \vec{u}_2 \cdots, \vec{u}_{N_s} \}$ with
$\vec{u}_i = ( u_{1 i}, u_{2 i}, \cdots, u_{N_\omega i} )^t$.
Then, one immediately observes from 
eqs.(\ref{br3}) and (\ref{br4})
that $\vec{a}$ is in the singular space. This implies that
$\vec{a}$ can be parametrized only by  
a set of $N_s$ parameters $(b_1, \cdot \cdot \cdot b_{N_s})$ as 
\beq
\label{br5}
  \vec{a} = \sum_{i=1}^{N_s} b_i \vec{u_i}, \ \ \ {\rm i.e.,}
 \ \ \  a_l = \sum_{i=1}^{N_s} U_{l i} b_i \ \ .
\eeq
 Therefore, owing to $U^tU=1$, eq.(\ref{br3}) reduces to 
\begin{eqnarray}
\label{br5-1}
-\alpha  \vec{b} = \vec{g} \equiv \Xi' \, V'^t 
\overrightarrow{  \frac{\partial L}{\partial D_A} } ,
\end{eqnarray}
where $\vec{b}$ is an $N_s$ dimensional column vector,
and $\Xi'$ and $V'$ are an $N_s \times N_s$ matrix and an
$N \times N_s $ matrix obtained by restricting
$\Xi$ and $V$ to the singular space, respectively.
 In other words, they are defined by
$\Xi'_{ij} = \Xi_{ij}$ $(i,j=1, 2, \cdots, N_s)$ and
$V'_{ij} = V_{ij}$ $(i=1, 2, \cdots, N,~~j=1,2, \cdots, N_s)$.

To solve (\ref{br5-1}), 
the standard Newton method is used for each increment
$\delta \vec{b}$ as
\begin{eqnarray}
\label{newton0}
J \ \delta \vec{b} = - \alpha \vec{b} - \vec{g},
 \ \ \ {\rm with } \ \ \ 
J_{ij} = \alpha  I_{ij} + \frac{\partial g_i}{\partial b_j} ,
\end{eqnarray}
where $I$ is the identity matrix.
By using the chain rule and the identity
${\partial A}/{\partial b} = \mbox{diag}[A] \, U $,
(\ref{newton0}) is rewritten as
\begin{eqnarray}
\label{newton1}
\left[\ (\alpha + \mu )\  I + MT \right]\
   \delta \vec{b} = -\alpha \vec{b} - \vec{g},
\end{eqnarray}
where
\begin{eqnarray}
M \equiv \Xi' \, V'^t \, 
 {\partial^2L \over \partial D^A \partial D^A } \, V' \, \Xi' , \ \ \  
  T \equiv U'^t \, \mbox{diag}[A] \, U',
\end{eqnarray} 
with   $U'$ being  defined by $U'_{ij} = U_{ij}$ 
$(i=1,2,\cdots, N_\omega, ~~j=1,2,\cdots, N_s)$.

Note   that 
the Marquardt-Levenberg parameter $\mu$  is added 
 to the diagonal element of the Jacobian matrix \cite{NR}, so
 that 
 the increment $\delta \vec{b}$ at each iteration becomes small
 enough to guarantee the validity of the lowest order approximation used
 in the Newton method eq.(\ref{newton0}).
 As $\mu$ increases, the increment $\delta \vec{b}$ generally decreases.
  The choice of $\mu$ is not unique; here we follow ref. \cite{bryan}
 and adjust $\mu$ 
 so that the norm of $\delta \vec{A}$ defined by 
 the metric  $g_{ll'} = 
  \left( 1/A_l \right) \delta_{ll'}$ (see 
   (\ref{pr8}) in Appendix A)
  does not exceed the integrated default model:
\beq
 {\delta
 \vec{A}}^{\ t} \mbox{diag}[1/{A_l}] \, \delta \vec{A} = {\delta
 \vec{b}}^{\ t} T \, \delta \vec{b} \leq 
 c  \sum_{l=1}^{N_\omega} \, m_l \ \ ,
\eeq
 with $c$ being a constant of $O(1)$.

  In our actual analysis, we use the SVD routine  in ref. 
 \cite{NR}. Since the elements of the kernel $K^t$
 vary many orders of magnitude,
  quadruple precision is necessary
  for obtaining reliable result of spectral functions at low frequencies.
  Then, at
  each iteration in (\ref{newton1}), we start with $\mu =0$ 
 and increase $\mu$ by 10 multiples of $10^{-4} \alpha$
 until the norm condition with  $c=0.2$ is satisfied.
 Once this condition is fulfilled, 
 the increment $\delta \vec{b}$ is added to the temporary solution vector
 $\vec{b}_{temp}$, $\vec{b}_{new} = \vec{b}_{temp} + \delta \vec{b}$. This
 process is iterated until $|\delta Q/Q| < 10^{-5}$ is achieved.
 
 Due to the correlation in the imaginary time direction in each
 Monte Carlo sample on the lattice, we must take into account the
 non-diagonal covariance matrix $C$ defined by eq. (\ref{covariance}).
 In this case, the Bryan method is applied after the following 
 transformations \cite{physrep,linden},
\begin{eqnarray}
K \to \bar{K} = R^{-1} K, \,\,\,\, D \to \bar{D} = R^{-1}D, 
\label{eq002}
\end{eqnarray}
where $R$ is a matrix which diagonalizes the covariance matrix,
$R^{-1} C R  = \mbox{diag} \, [ \, \bar{\sigma}_i^2 \,]$.
Since $C$ is a real symmetric matrix, we  take an orthogonal
matrix for $R$, i.e., $R^{-1} = R^t$.
$\bar{K}$ and $\bar{D}$, instead of $K$ and $D$, are taken as the kernel
and the data, respectively.
After the transformation, the likelihood function $L$ is written as  
\begin{eqnarray}
\label{like-rot}
L = {1 \over 2} \sum_{i=1}^{N} 
\  \left( { \bar{D}_i - \sum_{l=1}^{N_{\omega}} \bar{K}_{il} A_l } 
\right )^2 /
{\bar{\sigma}_i^2}.
\end{eqnarray}
Using this expression, we can carry out the Bryan
method as in the case where the covariance matrix is diagonal. 

\subsection{Uniqueness of the solution in MEM}

In order to show the uniqueness of the solution of (\ref{max1}),
we first prove the following proposition.

\noindent
{\bf Proposition}:\\
\noindent
Consider   a real and smooth function $F$ with $n$ real variables,
 namely  $F(x_1, x_2, \cdots, x_n) \in {\bf R} $ with 
$(x_1, x_2, \cdots, x_n) \in {\bf R}^n$.
Suppose the matrix $\partial ^2 F / \partial x_i \partial x_j$
is negative definite, i.e.,
\begin{equation}\label{assump_rolle}
\sum_{i,j = 1}^{n}y_i
\frac{\partial ^2 F}{\partial x_i \partial x_j} y_j < 0 ~~~
({\rm for~any~} y_i \in {\bf R}-\{ 0 \} ) ,
\end{equation}
then $F$ has only one maximum if it exists.
In other words, the solution of
$ \partial F /\partial x_i  = 0~~(i=1,2,\cdots,n)$
is unique if it exists.

\noindent
{\bf Proof}:\\
\noindent
Assume that  there are more than or equal to two solutions for 
$ \partial F/ \partial x_i  = 0~~(i=1,2,\cdots,n)$. 
Take any two of them $\vec{x}_1$ and $\vec{x}_2$, and 
 define an interpolation 
 $\vec{x}(t) = \vec{x}_1 + t(\vec{x}_2 - \vec{x}_1 )
\equiv \vec{x}_1 + t \vec{y}$,
and $G(t) \equiv F(\vec{x}(t))$.
 From the assumption,  $ dG(t) / dt$ is continuous and
differentiable in $[0,1]$, and satisfies
\begin{equation}
\left . \frac{dG}{dt}\right | _{t=0} = 
\left . \frac{dG}{dt}\right | _{t=1} =  0.
\end{equation}
 Thus, from Rolle's theorem, 
there exists at least one  $t \in (0,1)$ such that
\begin{equation}\label{rolle}
\frac{d^2 G(t)}{dt^2}
=   \sum_{i,j=1}^{n}
y_i \left . \frac{\partial^2 F}{\partial x_i  \partial x_j}
\right | _{\vec{x} = \vec{x}(t)} y_j 
=  0.
\end{equation}
 However,  (\ref{rolle})  
contradicts (\ref{assump_rolle}).
Hence  there cannot be more than or equal to two solutions for 
$ \partial F /\partial x_i  = 0~~(i=1,2,\cdots,n)$.
If there is a solution for
$ \partial F / \partial x_i  = 0~~(i=1,2,\cdots,n)$,
it is the global maximum of $F$ from eq.(\ref{assump_rolle}).
Thus the proposition is proved. (QED)

\vspace{0.5cm}

We now proceed to prove 
that   the solution of eq.(\ref{max1}) is unique
  and  corresponds to the global maximum 
of $Q=\alpha S - L$ if it exists.

For an arbitrary $N_\omega$ dimensional non-zero real-vector
$ \vec{z} = (z_1,z_2,\cdots,z_{N_\omega})$, $\alpha S$ satisfies that
\begin{equation}\label{hesse_s}
\sum_{l,l' = 1}^{N_\omega}
z_l \frac{\partial^2 (\alpha S)}{\partial A_l \partial A_{l'}} z_{l'}
= - \alpha \sum_{l=1}^{N_\omega} \frac{ z_l^2} {A_l} < 0,
\end{equation}
where we have used  $0 \le A_l < \infty$ and $0< \alpha < \infty $.
It is important to notice that the l.h.s. never becomes zero.

On the other hand, from (\ref{like-rot}),
\begin{equation}\label{hesse_l}
\sum_{l,l' = 1}^{N_\omega}
z_l \frac{\partial^2 (-L)}{\partial A_l \partial A_{l'}} z_{l'}
= - \sum_{i=1}^{N}\frac{\bar{z}_i^2 }{\bar{\sigma}_i^2} \le 0,
\ \ \ {\rm with} \ \ \ 
\bar{z}_i = \sum_{l=1}^{N_\omega}\bar{K}_{il} z_l .
\end{equation}
The l.h.s. of (\ref{hesse_l}) becomes zero
 in the direction where  $\bar{z}_i=0$ $(i=1, \cdot \cdot \cdot N)$.
  There are many such 
 directions because the rank of $\bar{K}$ is at most $N$, which is 
 much smaller than $N_{\omega}$ (the dimension of the vector $z_l$).  
 This means that $-L$ has a lot of flat directions, and there is no
  unique maximum of $-L$ as a function of $A_l$.
   
Once one  adds (\ref{hesse_s}) to (\ref{hesse_l}), 
 the solution of eq.(\ref{max1}) becomes unique if it exists,
 because of the 
 proposition just proved.
The existence of $\alpha S$ ($ \neq 0$) in $Q=\alpha S - L$ plays an 
  essential role 
  for this uniqueness.

\subsection{More on the covariance matrix}

 The  eigenvalue spectrum of the covariance matrix $C_{ij}$ 
 is known to show a pathological behavior when $N_{conf}$ is not
large enough compared to $N$.
In fact, 
it is reported in ref.\cite{physrep}
 that the eigenvalue spectrum displays a sharp
 break when $N_{conf} < N $, i.e., some eigenvalues are
 of similar magnitude, while the others are much smaller.
 This leads to the likelihood function $L$ in
 (\ref{like-rot}) extremely large.
 Thus, it has been empirically preferred to take
 $N_{conf} > 2N$.
 In this subsection, we shall show that the small eigenvalues 
 found  for  $N_{conf} < N $ in ref.\cite{physrep} 
 are actually exact zeros.

First, we rewrite the definition of the covariance matrix
(\ref{covariance}) as follows:
\begin{equation}
C_{ij} = \sum_{m=1}^{N_{conf}}C_{ij}^{m} =
\sum_{m=1}^{N_{conf}} d_i^m d_j^m ,
\end{equation}
where $C_{ij}^{m}$ is an $N\times N$ matrix defined for
the $m$-th gauge configuration, and $d_i^m \equiv 
[N_{conf}(N_{conf}-1)]^{-1/2} (D^m(\tau_i) - D(\tau_i))$.

Now, it is easy to show that ${\rm Rank}[C_{ij}^m]=1$, since
 each column of $C_{ij}^m$ reads
 $d_i^m (d_1^m, d_2^m,   \cdots , d_N^m )^t$
 and is thus proportional to the same vector $[\vec{d}^m]^t$.
 Let us consider the case when $N_{conf} < N$. 
 Since ${\rm Rank}[A+B] \le {\rm Rank}[A]+ {\rm Rank}[B]$
for arbitrary matrices, $A$ and $B$, we obtain
\begin{equation}
{\rm Rank}[C_{ij}] 
 \le \sum_{m=1}^{N_{conf}} {\rm Rank}[C_{ij}^m] = N_{conf} < N \ \ .
\end{equation}
As a result, the number of zero eigenvalues $N_{zero}$ satisfies
\begin{equation}
N_{zero} = N- {\rm Rank}[C_{ij}] \ge N - N_{conf} \ \ .
\end{equation}
Therefore, 
\begin{equation}\label{ncond}
N_{conf} \ge N
\end{equation} 
is a necessary condition
to avoid the pathological behavior of the eigenvalues of the
covariance matrix.

As we shall see more  in Section 5, 
a substantial number of sweeps for gauge configurations
are inserted between measurements in the Monte Carlo simulation on the lattice.
 This is for carrying out
measurements   with minimally correlated
gauge configurations.
Thus, the direction of each $C_{ij}^m$ is expected to be independent,
and in most cases eq.(\ref{ncond}) is also a sufficient condition.
 In order to stabilize the behavior of the eigenvalues, however,
a condition such as $N_{conf} \ge 2N$ may be imposed \cite{physrep}.
In our lattice simulations discussed in Section 5,
this condition is well-satisfied. We have also checked explicitly
  that
 our calculation is free of the pathological behavior and that the
fluctuations along the $N$ principal axes obtained by diagonalizing
the covariance matrix are well-approximated by Gaussian forms.

\newpage

\setcounter{equation}{0}
\setcounter{subsection}{0}

\section{Analysis with mock data}

To check the feasibility of
the MEM procedure and to see the dependence of the 
MEM image on the quality 
of the data, we made the following test using mock data.
 Key issues here are  whether MEM can detect
 sharp peaks and   flat continuum simultaneously.
 To study this, we consider two cases:

\noindent 
{\bf Schematic SPF}:  MEM using  mock data obtained from a
 schematic spectral function with  
 two Gaussian peaks; one is sharp and the other is broad.

\noindent
{\bf Realistic SPF}: MEM using data obtained from 
a realistic spectral function
  parametrized to reproduce the $e^+e^-$ annihilation
cross section into hadrons.

In these tests, we have
assumed $T=0$ and that the covariance matrix
$C$ is diagonal for simplicity. The non-diagonality of $C$, however, 
plays an important role in the case of the actual lattice QCD data.
 
 \vspace{0.5cm}

 The basic strategy of the analysis is summarized as follows.
\begin{enumerate}
\item[(i)] We start with an input  image of the form
$A_{in}(\omega) \equiv \omega^2 \rho_{in}(\omega)$.
 $\omega^2$ is a factor expected from the
 dimension of the meson operators ($n=2$ in 
 eq.(\ref{a-asympt})).
 Then, we calculate the mock data from eq.(\ref{KAB}) at $T=0$ as
\beq
\label{mock-const}
D_{in}(\tau_i) = \int_0^{\omega_{max}} K(\tau_i , \omega) 
\  A_{in}(\omega) \ d\omega \ \ ,
\eeq
 where  $\omega_{max}$ may be chosen arbitrary large, but 
 we simply take some reasonable number above which
 $\rho_{in}$ does not show appreciable variation.

\item[(ii)] By taking $D_{in}(\tau_i)$ at 
$N $ discrete points 
and adding a Gaussian noise, we create a mock data
$D_{mock}(\tau_i)$.
To mimic the noise level of our lattice QCD data with the 
 Dirichlet boundary condition in the temporal direction,
 the variance of the noise $\sigma (\tau_i)$
is chosen as
\beq
\label{noise-level}
\sigma (\tau_i )=
  b \times D_{in}(\tau_i) \times
 {\tau_i \over \Delta \tau } \ \ .
 \eeq
Here the  parameter $b$  controls the noise level.
 $N$ and $b$ are changed to see the sensitivity of the output
 image.

\item[(iii)] We construct an output image 
$A_{out}(0 \le \omega \le \omega_{max} ) = \omega^2 \rho_{out}(\omega)$
using MEM with the mock data $D_{mock}(\tau_i)$.
 Then, we  compare $A_{out}$ with $A_{in}$. We need to introduce
  the resolution $\Delta \omega$ in the frequency space to perform
  the numerical analysis in MEM as explained in the previous section;
   $\Delta \omega$ can be any small number. 
   As a possible measure for the quality of the output image, 
   we introduce a 
distance between $\rho_{out}$ and $\rho_{in}$ as
\beq
\label{distance}
r \equiv \int_0^{\omega_{max}} 
 \left[ \rho_{out}(\omega) - \rho_{in}(\omega) \right]^2 \ d\omega \ .
\eeq
\end{enumerate}

\subsection{Schematic SPF}

To study how the maximum entropy method can reproduce simultaneously
a sharp peak and a broad peak, 
we first consider the following schematic spectral function:
\beq
\rho_{in} (\omega) = \sum_{j=1,2} 
 {1 \over \sqrt{\pi} (\Gamma_j /2) } e^{-(\omega - M_j)^2 /(\Gamma_j /2)^2},
\eeq
with
$ (M_1,\Gamma_1) = (1, 0.01)$ GeV, $(M_2, \Gamma_2) = (3, 0.5) $ GeV.

To mimic the lattice data discussed later, 
the  parameters for the analysis here  are chosen to be 
$\omega_{max}=6$ GeV, $\Delta \omega=10$ MeV, $\tau_{min} =
\Delta \tau =0.085$ fm. To see 
how $A_{out}$ is improved as the quality of data 
 increases, $N =\tau_{max}/\Delta \tau$ 
and $b$ are changed within
the intervals, $10 \le N \le 30$  and $0.0001 \le b \le 0.01$. 
As for $m$, we take the form  $m(\omega) = m_0 \omega^2$, where
  $m_0$ is chosen to be 0.36 so that $\int_0^{\infty} \rho_{in}(\omega)
 \omega^2 d\omega = \int_0^{\infty} m_0 \omega^2 d\omega $ is satisfied.
 
  In Fig.\ref{fig1}, a comparison of $\rho_{out}(\omega)$ (the solid line) and 
  $\rho_{in}(\omega)$ (the dashed line) is shown for various combinations
   of $N$  and $b$.  The distance $r$ defined in (\ref{distance})
    is also shown in the figure.
 Increasing $N$ 
 and reducing the noise level $b$ lead to better
 SPFs closer to the input SPF as is evident from the figure.
 MEM reproduces not only the broad peak but also the sharp  peak without
 difficulty. This is because 
 the entropy density in (\ref{entropy}) is a local function
 of $\omega$ without any derivatives and hence does not introduce
 artificial smearing effect for SPF in the $\omega$-space.
 Fig.\ref{fig1} also shows  that, within the range of parameters varied,
 decreasing $b$ is more important than increasing $N$ to obtain
 a better image.  

 \begin{figure}[hbt]
\vspace*{0.5cm}
\hspace*{-0.5cm}
\epsfxsize=15cm
\centerline{\epsfbox{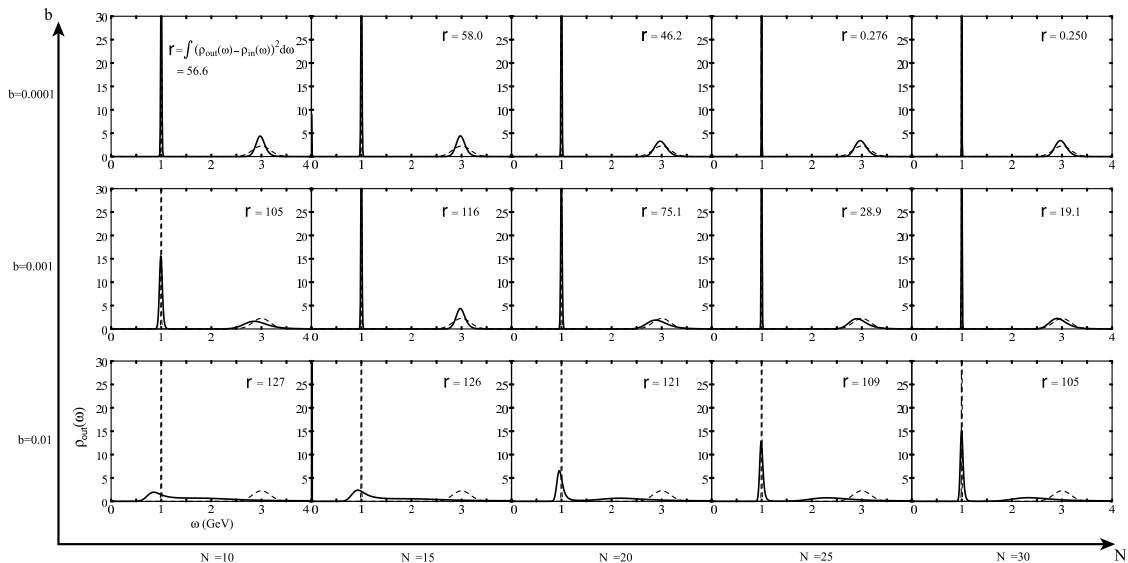}} 
\vskip2mm
\caption{Input SPF with two Gaussian peaks  (the dashed lines)
and output SPF obtained by MEM (the solid lines)
for different values of data-points $N$ and noise level $b$.}
\label{fig1}
\end{figure}

 In Fig.\ref{fig2}, the probability distribution $P[\alpha|DHm]$ 
 (with an approximation discussed in Step 2 in Section 3.3) used
  to obtain the final image is shown for three different combinations 
   of $N$ and $b$. The distribution tends to become 
  peaked as  $N$ increases and $b$ decreases.

\begin{figure}[hbt]
\hspace*{-0.5cm}
\epsfxsize=7cm
\centerline{\epsfbox{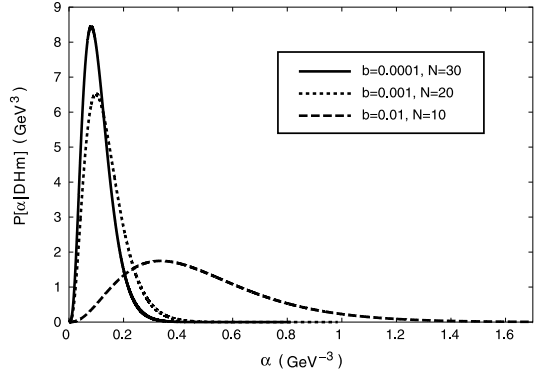}} 
\vskip2mm
\caption{ Probability distribution $P[\alpha|DHm]$ for
three different sets of $(N, b)$ in the case of   
 Schematic SPF. }
\label{fig2}
\end{figure}
 
 \subsection{Realistic SPF}

 As an example of  realistic spectral functions, we study
 SPF in the charged $\rho$-meson channel.  
 The isospin symmetry relates it to the $e^+e^-$ annihilation data
 into hadrons in the isospin 1 channel.
  We take a relativistic
Breit-Wigner form as a parametrization of 
 SPF in this channel \cite{shuryak}:
\beq
\label{realistic}
\rho_{in}(\omega) 
 =  {2 \over \pi} \left[
 {F_{\rho}^2 } {\Gamma_{\rho} m_{\rho} \over (\omega^2 - m_{\rho}^2)^2
 + \Gamma_{\rho}^2 m_{\rho}^2 } + {1 \over 8 \pi}
 \left (1+ {\alpha_s \over \pi } \right )
 {1 \over 1 + e^{(\omega_0 - \omega)/\delta}} \right] .
\eeq
The pole residue $F_{\rho}$ is defined:
\beq\label{fr}
 \langle 0 | \bar{d} \gamma_{\mu} u | \rho \rangle
 = \sqrt{2} F_{\rho} m_{\rho} \epsilon_{\mu} 
 \equiv   \sqrt{2} f_{\rho} m_{\rho}^2 \epsilon_{\mu} ,
\eeq
where $\epsilon_{\mu} $ is the polarization vector.
In the vector dominance model, the dimensionless 
 residue $f_{\rho}$ is related to the 
 $\rho \pi \pi$ coupling $g_{\rho \pi \pi} $ as
 $f_{\rho} = 1/g_{\rho \pi \pi} $.

 To get the correct threshold behavior due to the
 $\rho \rightarrow \pi \pi$ decay, we take the following
 energy-dependent width:
\beq
\Gamma_{\rho} (\omega) = {g_{\rho \pi \pi}^2 \over 48 \pi}
 \ m_{\rho} \left( 1- { 4m_{\pi}^2 \over \omega^2} \right)^{3/2}
 \ \theta(\omega - 2 m_{\pi}).
\eeq

The empirical  values of the parameters are 
\beq
m_{\rho} & = & 0.77 \ {\rm GeV}, \ \
m_{\pi}  =  0.14 \ {\rm GeV}, \nonumber \\
g_{\rho \pi \pi}  & = &  5.45, \\
\omega_0 & = & 1.3 \ {\rm GeV},   \ \ 
\delta =  0.2 \ {\rm GeV} \nonumber ,
\eeq
where we have assumed the vector dominance and 
$ \alpha_s $ is taken to be 0.3 
  independent of $\omega$ for simplicity. 

As for $m$, we take the form 
$m(\omega) = m_0 \omega^2$
motivated by the
asymptotic behavior of $\rho_{in}(\omega)$ in (\ref{realistic}).
$m_0$ is chosen to be 0.0257, which is slightly smaller than
0.0277 expected from the large $\omega$ limit of (\ref{realistic}), 
$\rho_{in} (\omega \rightarrow \infty ) = (2/\pi)(1/8\pi)(1+\alpha_s/\pi)$.
The same parameters as those for the schematic SPF are chosen:
$\omega_{max}=6$ GeV, $\Delta \omega=10$ MeV, $\tau_{min} 
= \Delta \tau =0.085$ fm.
$N =\tau_{max}/\Delta \tau$  and $b$ are changed within
the intervals, $ 10 \le N \le 30$  and $ 0.0001 \le b \le 0.01$.
 
In Fig.\ref{fig3}, the mock data  $D_{mock}(\tau_i)$ obtained from
(\ref{mock-const}) and (\ref{realistic}) are shown for the 
noise parameter $b = 0.001$. The linear slope of 
 ${\rm  log} D_{mock}(\tau_i)$ for
large $\tau_i$ is dictated by
 the lowest resonance, while the deviation from the linear slope for
   small $\tau_i$ originates from the continuum in SPF.  
 
\begin{figure}[hbt]
\hspace*{-0.5cm}
\epsfxsize=6cm
\centerline{\epsfbox{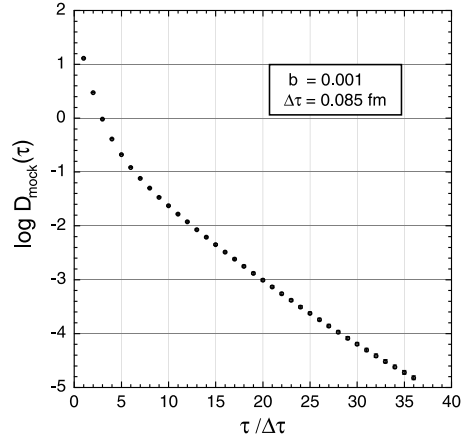}} 
\vskip2mm
\caption{Mock data $D_{mock}(\tau_i)$ in the case of the realistic SPF with
 Gaussian error attached.}
\label{fig3}
\vspace*{-0.3cm}
\end{figure}

\begin{figure}[hbt]
\vspace*{0.2cm}
\hspace*{-0.5cm}
\epsfxsize=15cm
\centerline{\epsfbox{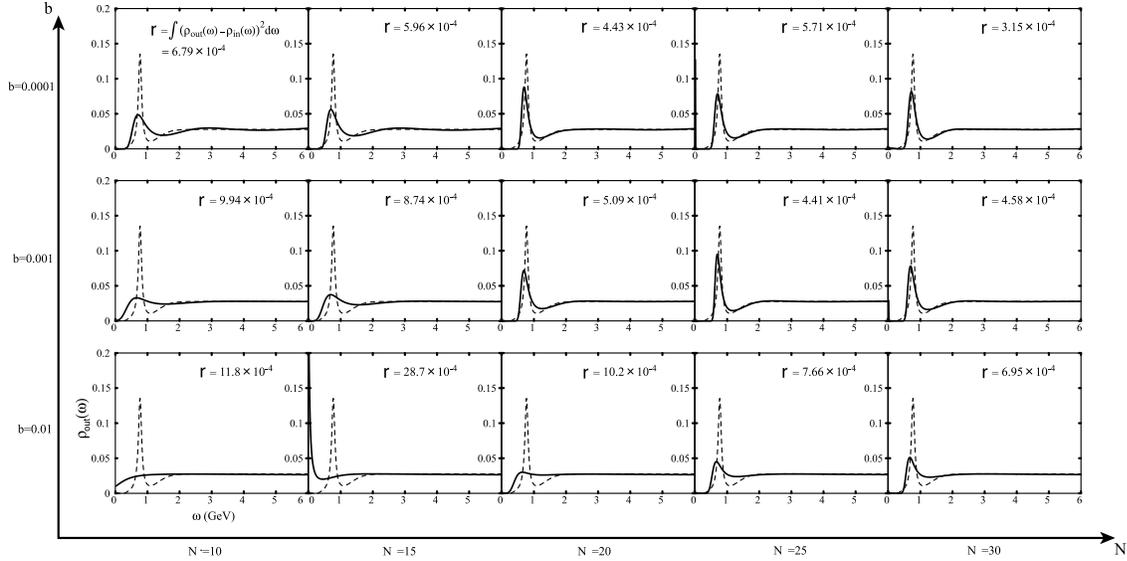}} 
\vskip2mm
\caption{Input  SPF with a resonance + continuum (the dashed lines)
 and output SPF  obtained by MEM (the solid lines) 
for different values of $N$ and $b$.}
\label{fig4}
\end{figure}

In Fig.\ref{fig4}, a comparison of $\rho_{out}(\omega)$ (the solid line) and 
$\rho_{in}(\omega)$ (the dashed line) is shown for various combinations
of $N$ and $b$. The distance $r$ defined in (\ref{distance})
is also shown in the figure.
Increasing $N$ 
and reducing the noise level $b$ lead to better
SPFs closer to the input SPF as is evident from the figure
as in the case of the schematic SPF.

In Fig.\ref{fig5}, the probability distribution $P[\alpha|DHm]$ 
  (with an approximation discussed in Step 2 in Section 3.3) used
  to obtain the final image is shown for three different combinations 
  of $N$ and $b$. The narrower distribution is
  obtained when the data quality is better.
  
\begin{figure}[hbt]
\vspace*{0.7cm}
\hspace*{-0.5cm}
\epsfxsize=7cm
\centerline{\epsfbox{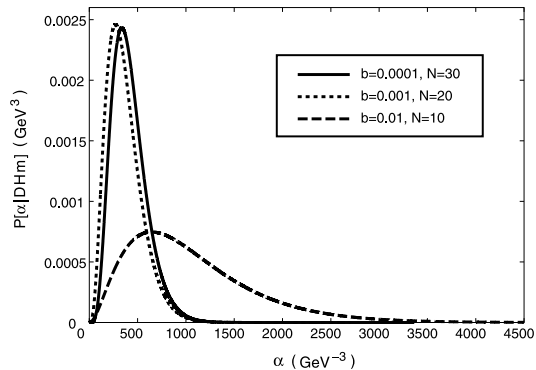}}

\vskip2mm
\caption{Probability distribution $P[\alpha|DHm]$ for
three different sets of $(N, b)$ in the case of the 
 realistic SPF. }
\label{fig5}
\end{figure}

\vspace*{1.5cm}
\begin{figure}[hbt]
\vspace*{-1.2cm}
\hspace*{-0.5cm}
\epsfxsize=15cm
\centerline{\epsfbox{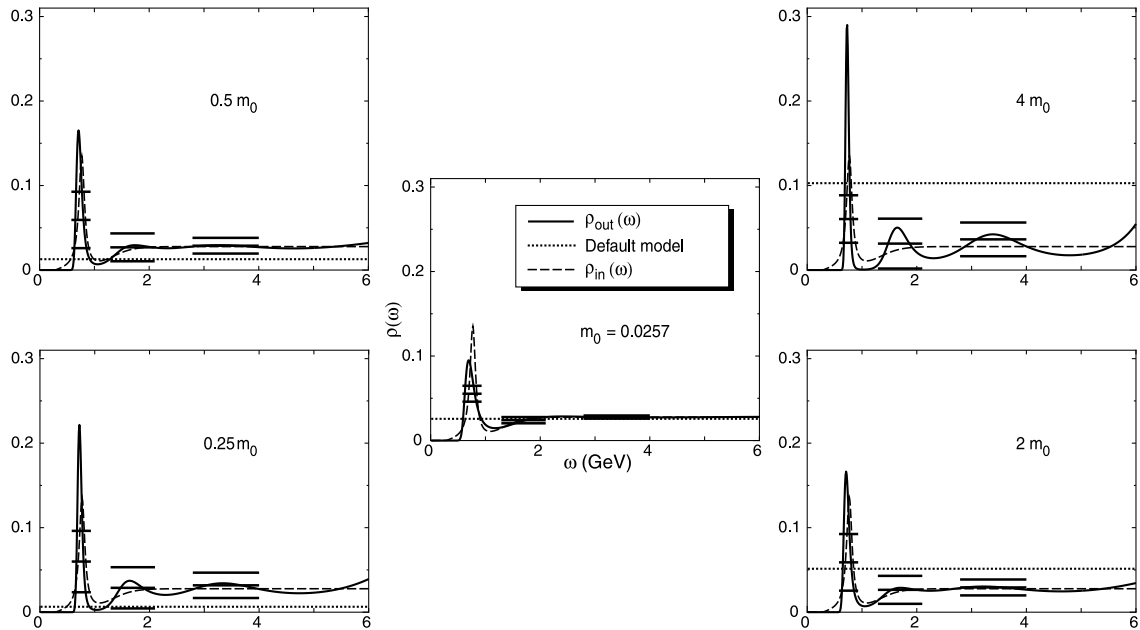}}

\vskip2mm
\caption{
Default model dependence of the output image $\rho_{out}(\omega)$
 (solid lines) for $N=25$ and $b=0.001$.
 The input images and the default models
 are shown by the dashed and dotted lines, respectively.
 Error bars for three different frequency regions are also
 attached. The middle figure is the same as that
 in Fig.\ref{fig4}.}
\label{fig6}
\end{figure}

\newpage

  Finally, to see the sensitivity of the results against the change of $m_0$,
 $\rho_{out}(\omega)$ is shown in Fig.\ref{fig6}
  for five different values of the default model together with
 the error bars.
 $N=25$ and $b=0.001$ are chosen.
   Short dashed and long dashed lines correspond to 
 the default model and the input SPF, respectively.
 The solid lines show the output image after the MEM analysis.
 $m_0 = 0.257$ is chosen in the
 middle figure  which is the same with that in 
 Fig.\ref{fig4}, while the other four figures
 correspond to the default models for $0.25 m_0$, $0.5m_0$, $2m_0$
 and $4 m_0$. Even under the factor 4 variation of the
 default model, the resultant SPFs show the peak + continuum structure.
 However, as the default model deviates from the expected asymptotic
 value, the SPF starts to have a ``ringing'' behavior.

 Now, the error analysis discussed in Step 3 in Section 3.3 can tell
   us whether the ringing structure seen, e.g., in  
   the $4m_0$ case (the upper right figure)
  corresponds to a  real resonance or is the artifact of the 
 maximum entropy method.
  The horizontal position and length of the 
 bars in Fig.\ref{fig6} indicate the frequency region over which 
 the SPF is averaged, while the vertical height of 
 the bars denotes the uncertainty in the averaged value
 of the SPF in the interval.
 Implications of the error bars in Fig.\ref{fig6}
 are twofold;  (i) the ringing images for
 $4 m_0$ and $0.25 m_0$ are statistically not significant, 
 and (ii) the combination of the best SPF and the 
 best default model may be selected  by
 estimating the error bars (Step 4 in Section 3.3).
 In the present case,
 for the given data, the middle figure should be chosen to be
 the best one.

\newpage

\setcounter{equation}{0}
\setcounter{subsection}{0}

\section{Analysis with lattice QCD data}

\subsection{Lattice basics}

In lattice gauge theories \cite{lattice_general}, 
the Euclidean space-time is
discretized into cells. In our
simulations, we use an isotropic hypercubic grid,
 namely the lattice spacing $a$ is the same in all directions.
 In the following, we assume that the gauge group is SU(3)
with QCD in mind. The fermion
field  $\psi(x)$  is defined at grid sites, while
  the gluon field is defined on links 
  connecting adjacent pairs of grid sites (Fig.\ref{fig7}).
On the  links connecting  $x$ to 
$x\pm \hat{\mu}$  ($|\hat{\mu}| = a$),
we define $U_{\pm \mu}(x)$ as 
\begin{eqnarray}
U_{\mu}(x) =  \exp\left
[ iag A_\mu   (x    ) \right ], \ \ \ 
U_{-\mu}(x)  = \exp\left
[ -iag A_\mu   (x -  \hat{\mu}   ) \right ]. 
\end{eqnarray}
$U_{\mu}(x)$ is an element of
the SU(3) group,
$g$ is the gauge coupling constant, and $A_\mu$ is
the gauge field.
The local gauge transformation for $\psi(x)$ and
$U_{\mu}(x)$ on the lattice reads
\begin{eqnarray}
\psi(x) & \rightarrow & V(x)\psi(x), \ \ \ 
\bar{\psi}(x)   \rightarrow  \bar{\psi}(x)V^{\dag} (x), \nonumber \\
U_\mu (x) & \rightarrow & V(x+\hat{\mu}) U_\mu (x) V^{\dag} (x),
\label{lat_gaugetr}
\end{eqnarray}
where $V(x)$ is an SU(3) matrix defined at each site $x$.

\begin{figure}[hbt]
\vspace*{0.5cm}
\hspace*{-0.5cm}
\epsfxsize=5cm
\centerline{\epsfbox{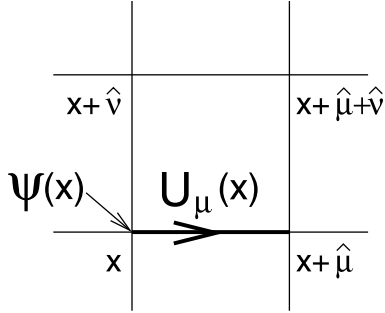}}

\vskip2mm
\caption{Lattice variables. Quarks (gluons) are defined
 at the sites (on the links). }
\label{fig7}
\vspace*{0.3cm}
\end{figure}

Using the plaquette variable defined by
\begin{equation}
W(x,\hat{\mu},\hat{\nu}) =   
U_{-\nu} (x  +\hat{\nu})U_{-\mu} (x  +\hat{\mu}+\hat{\nu})
U_\nu (x+\hat{\mu})U_\mu (x),
\end{equation}
the single  plaquette gluon action is written as 
   \begin{eqnarray}
S_p   =  \beta_{lat}  \sum_{x}\sum_{\mu < \nu}
{\rm Re}\,{\rm Tr}\,\frac{1}{3}
\left (1 - W(x,\hat{\mu},\hat{\nu}) \right )  ,\ \ \ {\rm with} \ \ 
\beta_{lat} = {6 \over g^2}.
\end{eqnarray}
In the naive continuum limit $a\rightarrow 0$, $S_p$ 
approaches the continuum gauge action
\begin{equation}
S_{cont} = \frac{1}{4} \int d^4x (F_{\mu \nu}^a(x))^2.
\end{equation}

For fermions, we use the Wilson quark action $S_w$  defined by
\begin{eqnarray}
S_w & = & \sum_{x,y}\bar{\psi}(x)
\left [ \delta_{xy} - \kappa\sum_{{\mu}}
\delta_{x,y+\hat{\mu}}(1 + \gamma_{\mu}) U_{ \mu } (y)
\right ]
\psi(y) \nonumber \\
& \equiv &  \sum_{x,y}\bar{\psi}(x) D_w(x,y;U) \psi(y).
\label{defq}
\end{eqnarray}
Here the flavor indices for fermions are suppressed. 
$\gamma_{\mu} $ satisfies $\{ \gamma_{\mu}, \gamma_{\nu} \} 
 = 2 \delta_{\mu \nu}$. $\Sigma_{\mu}$ is taken for both positive and 
 negative directions of $\mu$ together with a convention
 $\gamma_{-\mu} = - \gamma_{\mu}$.
 $\kappa$ denotes the hopping parameter, while the quark mass $m_q$ is
defined as 
\begin{equation}
m_q a = \frac{1}{2\kappa} - \frac{1}{2\kappa_c(\beta_{lat})},
\end{equation}
for  a fixed $\beta_{lat}$.
Here $\kappa_c(\beta_{lat})$ 
is the critical hopping parameter, at which the pion becomes massless.

The action for the whole system $S_{lat}$ is given by the
sum of $S_p$ and $S_w$,
\begin{equation}
S_{lat} = S_p + S_w .
\end{equation}
The expectation value of a physical observable
${\cal O}[U, \psi, \bar{\psi}]$
is given in terms of path integrals over
the link variables $U_\mu (x)$ and fermion fields  
$\psi(x)$ and $\bar\psi(x)$, 
\begin{equation}
\langle {\cal O}[U, \psi, \bar{\psi}] \rangle 
= \frac{1}{Z}\int [dUd\bar{\psi} d\psi]\ 
e^{-S_{lat}} \ {\cal O}[U, \psi, \bar{\psi}],
\end{equation}
where $Z$ is given by
\begin{equation}
Z = \int [dU d\bar{\psi} d\psi]
e^{-S_{lat}}  = \int [dU] e^{-S_p}\ {\rm det}D_w( x,y;U).
\end{equation}
In the quenched approximation, ${\rm det}D_w( x,y;U)$ is set to 1.
 Physically, this corresponds to ignoring the effect
of virtual quark loops. This approximation simplifies numerical
simulations by factor 1000 or so, since it is very time consuming to
calculate the determinant of a huge matrix such as  $D_w( x,y;U)$.
In the quenched simulations, the first step is to 
 generate   an ensemble of gauge configurations $U_\mu (x)$ 
  with the weight $e^{-S_p}$.
A typical method for that purpose 
in SU(3) gauge theory is  
 the  pseudo heat-bath method combined with the  over-relaxation
method (for more details, see, e.g., \cite{creutz_ph}).
 
 \subsection{Lattice parameters}
 
 We take the open MILC code \cite{milc} with minor modifications  and
 perform simulations with 
 the single plaquette gluon action
 + Wilson quark action in the quenched approximation
 on  a Hitachi 
 SR2201 parallel computer at Japan Atomic Energy Research Institute.
 The basic lattice parameters in our simulations are
\beq
{\rm coupling \  strength} 
& &  \beta_{lat}  =  {6 \over g^2} = 6.0, \nonumber \\
{\rm lattice \ size}
 & &  N_{\sigma}^3 \times N_{\tau}  =  20^3 \times 24, \nonumber \\
{\rm hopping \ parameter} & & 
 \kappa = 0.153,\  0.1545\  {\rm and}\  0.1557.
\eeq
 The Dirichlet boundary condition (DB)
 in the temporal direction, which is defined by
 $U_{\mu}(x)=0$ for the last temporal links, 
 is employed to have
 as many temporal data points as possible.
 In the spatial directions,
 the periodic (anti-periodic)  boundary condition is used for the
 gluon (quark) as usual.
 Gauge configurations are generated by the pseudo heat-bath and
 over-relaxation algorithms with a ratio $1:4$. Each configuration
 is separated by 200 sweeps. The number of gauge configurations 
 used in our analysis is $N_{conf}=161$.

 We have also carried out simulations with the periodic
 (anti-periodic) boundary condition for the gluon (quark)
 in the temporal direction on the $20^3 \times 24$ lattice 
 with $\beta_{lat} = 6.0$, and on the $40^3 \times 30$ lattice
 with $\beta_{lat} = 6.47$ on CP-PACS at Univ. of Tsukuba.
 Detailed MEM study in those cases will be reported elsewhere
 \cite{ahn_finer}.
    
To calculate the two-point correlation functions,
we adopt a point-source at $(\tau, \vec{x})=(0,\vec{0})$,
 and a point-sink at time $\tau$ with the spatial points
 averaged over the spatial lattice 
 to extract  physical states with vanishing three-momentum 
 (see eq. (\ref{KAB})). 
 The following local and flavor non-singlet operators are adopted 
 for the simulations:
\begin{eqnarray}
\label{operators}
 & & {\rm scalar\  (S)}:\  \bar{d}u,
 \ \ \ \ \ \  \ \ \ 
 {\mbox{pseudo-scalar} \  ({\rm PS})}:\   \bar{d} \gamma_5 u, \\
\label{operators2}
 & & {\rm vector\  (V)}:\  \bar{d} \gamma_{\mu} u,
 \ \ \ \ \ 
 {\mbox{axial-vector} \ ({\rm AV})}: \  \bar{d} \gamma_{\mu} \gamma_5 u.
\end{eqnarray}
In the V and AV channels, the spin average is taken 
 over the directions ${\mu} = 1,2,3$ to increase statistics.
Simulations for spin 1/2 and 3/2 baryons
 have been also carried out.
 Since special considerations are necessary for 
 the decomposition of the SPFs in the baryon channels,
 we shall report the results in a separate  publication \cite{ahn_baryon}.

 Note here that the use of the point source and sink is essential
 for obtaining good signals for the resonance and continuum
 in the spectral function simultaneously. First of all, 
 the SPF defined with the point source and sink
 for the vector channel is directly
 related to the experimental observables such as the 
 $R$-ratio and the dilepton production rate as discussed
 in Section 2.  Besides, there are two-fold disadvantages 
 to use smeared sources and sinks in the MEM analysis:
 First of all, it is difficult to write down a simple sum rule such as
 eq.(\ref{KAB}) for the smeared operators.
 Secondly, the coupling to the excited hadrons  becomes small
 for  such operators and one loses information on the 
 higher resonances and continuum.

 For the temporal data points to be used 
 in the MEM analysis, we take
 $\tau_{min}/a =1$ and  $\tau_{max}/a=12$.
 The latter is chosen to suppress the 
 error from the Dirichlet boundary condition. 
 In fact, we found that the statistical error of our data is
well parametrized  by formula (\ref{noise-level})
with a slope parameter $b$ up to $\tau/a = 12$ and
that the error starts to increase more rapidly above
$\tau/a =12$.

\subsection{Spectral functions and their asymptotic forms}

We introduce  the dimensionless SPFs, $\rho(\omega)$, as follows:
\begin{equation}
\label{def-spf2}
 A_{_{S,PS} }(\omega) = \omega^2 \rho_{_{S,PS} }(\omega) , \ \ \ \ 
 {1 \over 3} \sum_{\mu=1,2,3} A^{\mu \mu}_{_{V,AV}}(\omega) 
 = \omega^2 \rho_{_{V, AV} }(\omega) .
\end{equation}
$\rho_{_{S,PS,V,AV}}(\omega ) $ are defined in 
such a way that they approach
finite constants when 
$\omega \rightarrow \infty$ in the continuum limit ($a \rightarrow 0$)
as predicted by  perturbative QCD (see eq.(\ref{a-asympt})).

Because of  (\ref{sym1}) and (\ref{sym2}), we have 
\begin{eqnarray}
\rho_{_{S,PS,V,AV} }(\omega ) = - \rho_{_{S,PS,V,AV} }(- \omega ) \ge 0,
\ \ \ \ {\rm with}~~\omega\ge 0.
\end{eqnarray}
The SPFs on the lattice in the chiral limit should have the following 
 asymptotic form when $\omega$ is large and $a$ is small;
\begin{equation}
\label{asympt-form}
 \rho_{j}(\omega \gg {\rm 1~GeV})
 = { r_{1j} \over 4 \pi^2} \left ( 1 + r_{2j} 
  {\alpha_s(\mu ) \over \pi} \right )
 \left ( {1 \over 2\kappa_c  Z_{j}(\mu a) } \right )^2 ,
\label{cont-V}
\end{equation}
where $j={\rm S,\ PS, \ V {\rm and} \ AV}$.
The first two factors on the r.h.s. are the $q \bar{q}$ + $q\bar{q}g$
continuum expected from  perturbative QCD.
The third factor contains  the non-perturbative
renormalization constant, $Z_j(\mu a)$, of the lattice composite operator.
The renormalization point $\mu $ should be chosen to be the typical scale
of the system such as $1/a$.

 $r_{1j}$ and $r_{2j}$  calculated perturbatively in the
 continuum QCD \cite{RRY} are shown in Table \ref{cont.par}.
 In \cite{shi}, $Z_j(\mu a =1)$ in the chiral limit has been 
  calculated  numerically on the lattice
 with $\beta_{lat}=6.0$  for PB (the periodic (anti-periodic) boundary
  condition in the temporal direction for the gluon (quark)), which is 
  summarized in Table \ref{cont.par}.
   $Z_j$ for DB 
   and those for  PB are in principle different.
 This is because, in our simulation, the hadronic source is always on a 
 time slice  where DB is imposed and $Z_j$ detects  the 
 the boundary effect.    Our  measurement of the 
 decay constant  $f_{\rho}$  confirms this as shown later
 in  section 5.5.1. Therefore, we use 
 $Z_j$ listed in Table \ref{cont.par} only to guide our default model $m$. 

\begin{table}[ht]
 \begin{center}
  \begin{tabular}{|c|c|c|c|c|}
 \hline  
      $j$ & $r_{1j} $  &  $r_{2j}$  &  $Z_j(\mu a=1)$  
   &  $\rho_j (\omega$=$\mu$=$1/a$) \\
 \hline
   S      &  ${3 / 2}$  & ${11 / 3}$  &  0.77   & 0.80 \\
  PS      &  ${3 / 2}$  & ${11 / 3}$  &  0.49   & 2.00 \\
   V      &  $1          $  & $1$             &  0.68   & 0.59 \\
  AV      &  $1          $  & $1$             &  0.78   & 0.45 \\
 \hline
  \end{tabular}
 \end{center}
 \caption{
 Constants for spectral functions in the asymptotic region.
 $r_{1j}$ and $r_{2j}$ are taken from \cite{RRY}. 
  $Z_j (\mu a =1) $ for
 $\beta_{lat} = 6.0$ in the chiral limit
  $\kappa_c = 0.1572 $ is taken from  \cite{shi},
 where PB
 is used in the temporal direction.
 For the estimates in the last column,
 we use $\alpha_s (\mu \simeq 2 {\rm ~GeV})=0.3$
 \cite{PDG}, $\kappa_c = 0.1572$ and $a^{-1} \sim ~2$ GeV
 for $\beta_{lat} =6.0$. }
 \label{cont.par}
\end{table}

 So far we have assumed that the spatial momentum ${\bf k}$ 
 vanishes
 due to the spatial integration on the lattice.
 However, it is not exactly the case for any 
 finite set of gauge configurations.
 In most cases, the finite ${\bf k}$ error is harmless as far
 as ${\bf k}$ is small enough. However, it is potentially dangerous 
 in the AV channel, where the SPF
 for finite $k^{\mu} = (\omega, {\bf k})$ is written as
\beq
 A_{{\mu}{\nu}}(\omega, {\bf k})
 = (k_{\mu}k_{\nu} - k^2 g_{\mu \nu})\  \rho_{_T}(\omega,{\bf k})
    + k_{\mu}k_{\nu}\ \rho_{_L}(\omega,{\bf k}).
\eeq
 Here $\rho_{_T}$ is the transverse spectral function, which does not 
 have contamination from  the pion pole. When ${\bf k}=0$, $\rho_{_T}$
 reduces
 to $\rho_{_{AV}}(\omega \ge 0) \ge 0$ defined in (\ref{def-spf2}).
 On the other hand,
 $\rho_{_L}$ is the longitudinal spectral function, which contains 
 the pion pole: $\rho_{_L}(\omega,{\bf k}) = 2 f_{\pi}^2
 \delta(k^2-m_{\pi}^2) + \cdot \cdot \cdot $
 with $f_{\pi}$ being the pion decay constant.
 Then, if small amount of ${\bf k}$ remains,
 there is possible contamination from
 the low-mass pion pole to  the spin-averaged SPF:
 \beq
 {1 \over 3}\sum_{\mu=1,2,3}A_{\mu \nu}(\omega, {\bf k} \simeq 0)
 \simeq  \omega^2 \rho_{_{AV}}(\omega) + {\bf k}^2 \rho_{_L}(\omega).
 \eeq
 In the actual lattice data, this effect appears  
 in $D({\tau})$ for large $\tau$.  A possible way to
 subtract the contamination of  $\rho_{_L}$ is to carry out the measurement
 with finite ${\bf k}$.
   
 \subsection{Discretization in $\omega$ space}

 In the MEM analysis,
 we need to discretize the $\omega$-space into pixels of an equal size
 $\Delta \omega$ and carry out the integration, (\ref{KAB}),
 approximately.
 The upper limit of $\Delta \omega$ is determined by the requirement
 that the discretization  error of the kernel $K$
 in (\ref{KAB}) is small enough, namely,
 $ \tau \cdot  \Delta \omega \ll 1 $, which reads
\begin{eqnarray}
\label{upper}
 \Delta \omega \ll {1 \over \tau_{max} } < {1 \over \tau }.
\end{eqnarray}
Thus $\tau_{max}^{-1} = (N_{\tau} a)^{-1} = 
(24 a)^{-1} \sim 90 {\rm MeV}$ is the upper
 bound of $\Delta \omega$ on our lattice. 

 Also, to obtain the good resolution of SPF, one needs
 to have small enough $\Delta \omega$ so that
 $|A_{l+1} - A_{l}|/|A_{l+1} + A_{l}| \ll 1$.
 In the maximum entropy method with singular value
 decomposition, there is no  problem in choosing
 an arbitrary small value for $\Delta \omega$.
 In fact, as we have explained,
 the maximum search of $Q$ is always limited in the
 $N_s$ dimensional singular space, and  $N_s$ is independent of
 $N_{\omega}$. Therefore, increasing $N_{\omega}$ or decreasing
 $\Delta \omega$ does not cause any numerical difficulty.
 In the following, we adopt $\Delta \omega = 9.5$ MeV, which satisfies
 (\ref{upper}) and  simultaneously 
 gives a good resolution to detect sharp peaks in the spectral function. 

Aside from $\Delta \omega$, we also need to choose
the upper limit for the $\omega$ integration, $\omega_{max}$.
Since this quantity should be comparable to or larger than the maximum
available momentum on the lattice, we choose
$\omega_{max} = 7.1~{\rm GeV} >  \pi /a \sim 6.9$ GeV.
We have checked that larger
values of $\omega_{max} $ do not change 
the result of $A(\omega)$ substantially, while smaller
values of $\omega_{max} $ distort the high energy end of the
spectrum. The dimension
of the image to be reconstructed  is $N_{\omega} \equiv
\omega_{max}/\Delta \omega \sim 750$,
which is in fact much larger than the maximum number of 
data points, $N = 24$, available on our lattice.

\subsection{Results of MEM analysis}

\subsubsection{Pseudo-scalar and vector channels}

Let us first consider the PS and V
 channels. The lattice data   $D(\tau_i)$ in these channels are shown in 
 Fig.\ref{fig8}.
 For the MEM analysis, we take the formula (\ref{KAB}) with 
 $\beta = 1/T = \infty$, since we use DB (the Dirichlet boundary condition).
 Also, $\tau_{min}/a =1$ has been taken to utilize the 
 information available on the lattice as much as possible, while
 $\tau_{max}/a=12$ has been chosen to
 suppress errors caused by DB as discussed
 in the previous section.

 Shown in Figs.\ref{fig9}(a) and \ref{fig9}(b) are the reconstructed SPFs 
 in these channels for different values of 
 $\kappa$. 
  $\omega_{max}$ is taken to be 7.1 GeV and
 $\Delta \omega = 9.5$ MeV.  We have used $m = m_0 \omega^2$ with
 $m_0 = 2.0~(0.6)$ for the PS (V) channel motivated by
 the perturbative estimate $m_0 \sim \rho(\omega \gg 1 {\rm ~GeV})$
 in eq.(\ref{cont-V}) with Table \ref{cont.par}.
 The sensitivity of the reconstructed SPFs on the variation of
 $m_0$ will be discussed later.

 Spectral functions
  thus obtained in the PS and V channels shown in Figs.\ref{fig9}a
 and \ref{fig9}b
 have a common structure:
 a sharp peak at low-energy,
 a less pronounced second peak,
 and a broad bump at  high-energy. Let us discuss 
 those structures in detail below.

\begin{figure}[hbt]
\vspace*{0.5cm}
\hspace*{-0.5cm}
\epsfxsize=6cm
\centerline{\epsfbox{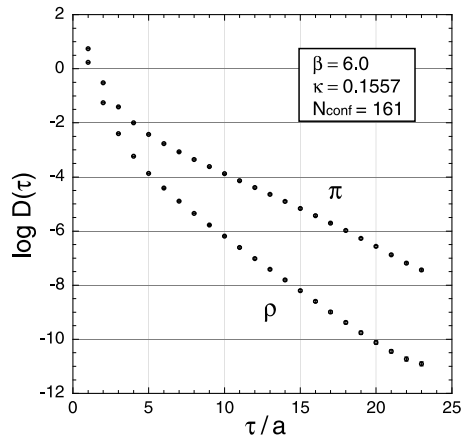}}

\vskip2mm
\caption{
	 Lattice data $D(\tau_i)$ in the PS and V channels with statistical
	 errors.}
\label{fig8}
\vspace*{-0.3cm}
\end{figure}
 
\begin{figure}[htb]
\vspace*{0.5cm}
\hspace*{-0.5cm}
\epsfxsize=12cm
\centerline{\epsfbox{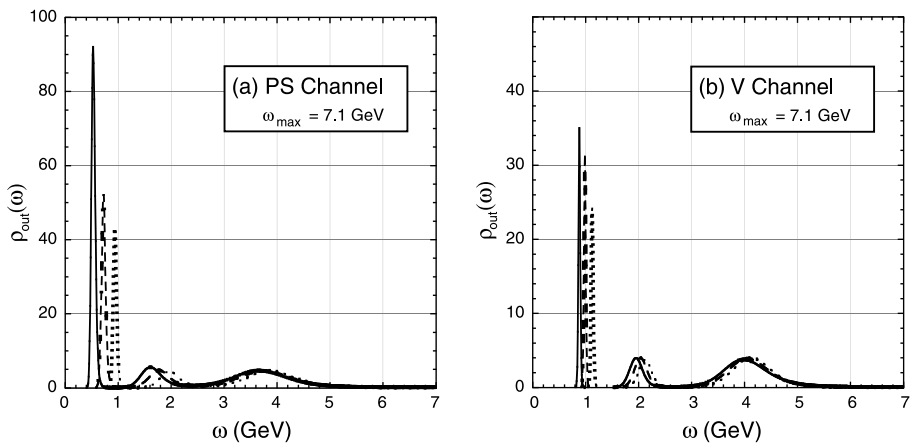}}

\vskip2mm
\caption{Reconstructed image $\rho_{out}(\omega)$ for
the PS (a) and V (b) channels. The solid, dashed and dash-dotted
lines are for $\kappa=$ 0.1557, 0.1545 and 0.153, respectively.
For the PS (V) channel, $m_0$ is taken
to be 2.0 (0.60). $\omega_{max}$ is 7.1 GeV.
}
\label{fig9}
\end{figure}

 \begin{enumerate}

\item[(i)] To make sure that the lowest peak  for 
 each $\kappa$ 
 in the PS (V) channel corresponds to the 
 pion (the $\rho$-meson),
 we extract $m_{\pi}$ and $m_{\rho}$ in the chiral limit with
 the following procedure. 
 First we fit  the lowest peak by a Gaussian form and extract
 the position of the peak. 
 Then, the linear chiral extrapolation is made 
 for  $(m_{\pi}a)^2$ and $m_{\rho}a$.
 The results together with $a$ using the 
 input $m_{\rho}=0.77$ GeV are summarized 
 in the column ``MEM continuum kernel'' in Table \ref{mass.sum2}.
 They are consistent with those
 determined from the standard analysis using the asymptotic behavior
 of $D(\tau_i)$ in the interval $10 \le \tau_i/a \le 15$, which
 are shown in Table \ref{mass.sum2}
 in the column ``asymptotic analysis''.
 Also, our results are consistent with those from 
 the asymptotic behavior of $D(\tau)$ obtained by the QCDPAX
 Collaboration 
 on a $24^3 \times 54$ lattice with $\beta_{lat}=6.0$  \cite{kc}.
 They give $ \kappa_c = 0.1571 $, 
 $m_{\rho}a = 0.331(22) $ and $a^{-1} = 2.33 (15)$ GeV.

 Although it is quite certain that the lowest peaks for each $\kappa$
 in Figs.\ref{fig9}(a) and \ref{fig9}(b) 
 correspond to $\pi$ and $\rho$, the widths of these peaks
 in the quenched approximation do not have physical meaning, since
 they are artifact caused by the incompleteness of the information
 contained in lattice data.  
 Nevertheless, the integrated strength of the peak corresponds
 to the physical decay constant of the mesons.
 For example, the dimensionless decay constant  
 $f_{\rho}$ defined in eq.(\ref{fr})
 is related to the lattice matrix element as
\beq
\langle 0 | (\bar{d} \gamma_{\mu} u)_{lat} | \rho ({\bf k}=0) \rangle
 = \sqrt{2} \epsilon_{\mu} f_{\rho} m_{\rho}^2 
  {1 \over 2 \kappa Z_V} .
\eeq
Therefore, it is further related to the spectral integral
 near the resonance as
\beq
\int_{pole} \rho_{_V} (\omega) d\omega^2 = 2 f_{\rho}^2 m_{\rho}^2
\  \left( {1 \over 2 \kappa Z_V} \right)^2,
\eeq
where the suffix ``$pole$'' implies the integral over the 
 lowest isolated peak. This integral can be carried out 
 numerically for each $\kappa$, and the linear extrapolation to the
 chiral limit has been done.
 The result for $f_{\rho}/Z_V$ 
 is given in the 2nd row of Table \ref{mass.sum2} together
 with that obtained
 from the asymptotic analysis. The agreement is satisfactory.
 As we have mentioned before, $Z_V$ in the DB case is not necessary
 equal to that in the PB case.
 Therefore, we cannot predict $f_{\rho}$ from our data alone.
 Instead, we
 extract $Z_V$ in the DB case by comparing
 our measurement of $f_{\rho}/Z_V$ with experimental value of $f_{\rho}$.
 This leads to
 $Z_{V}^{DB} \simeq 0.38 < Z_{V}^{PB}\simeq 0.68$.
 If we can have larger lattice and can place 
 the hadronic source and sink far from the boundary,
 the difference between $Z_j^{DB}$ and $Z_j^{PB}$ should 
 become small.

\item[(ii)] 
As for the second peaks in the PS and V channels,
the results of the 
error analysis, which are shown in Fig.\ref{fig10}, indicate that
their spectral ``shape" does not have much  statistical
significance, although the existence of the
non-vanishing spectral strength is significant.
With this reservation, we fit the position of the 
second peaks and make linear
extrapolation to the chiral limit. The results 
are summarized in the 
 3rd row of Table \ref{mass.sum2} together
 with the experimental values.
 Since 
 there exist two excited-$\rho$ experimentally observed,
  $\rho(1450)$ and $\rho(1700)$, 
 we quote both values for 
  $m_{\rho'}/m_{\rho}$ in Table \ref{mass.sum2}.

One should note here that, 
in the standard two-mass fit of $D(\tau)$, the mass
of the second resonance is highly sensitive to the 
lower limit of the fitting range, e.g., 
$m^{2nd}/m_{\rho} = 2.21(27)~(1.58(26))$  for $\tau_{min}/a = 8~(9)$
in the V channel with $\beta_{lat}=6.0$ \cite{kc}.
This is because the contamination from the short distance
contributions, which dominates the correlators at 
$\tau < \tau_{min}$, is not
under control in such an approach.
On the other hand,  MEM does not suffer from this difficulty 
and can utilize the full
information down to $\tau_{min}/a=1$.
Therefore, MEM opens up
a possibility of systematic study of higher resonances 
with lattice QCD.

\item[(iii)]
As for the third bumps in Fig.\ref{fig9}, the spectral ``shape" 
is statistically not significant as is discussed in
conjunction with Fig.\ref{fig10}.
They should rather be considered to be a part of the
perturbative continuum instead of a single resonance.
Fig.\ref{fig9}  and Fig.\ref{fig12} given later    
show  that SPF decreases substantially
above 6 GeV, namely MEM automatically detects
the existence of the momentum cutoff on the lattice $\sim \pi/a$.
It is expected that MEM with the data on finer lattices leads to larger
ultraviolet cut-offs in the spectra.
Our preliminary analysis on a finer lattice
($40^3 \times 30$ lattice with  $\beta_{lat} = 6.47$)
in fact indicates that this is the case \cite{ahn_finer}.
 
\end{enumerate}

\vspace*{-0.5cm}
\begin{table}[hbt]
 \begin{center}
  \begin{tabular}{|c|c|c|c|c|}
   \hline
 & asymptotic analysis & MEM   & MEM &  Experiments \\
 &  & continuum kernel   &  lattice kernel &  \\
 & ($10 \le \tau_i/a \le 15$) & ($1 \le \tau_i/a \le 12$)   &  ($1 \le \tau_i/a \le 12$) &   \\
   \hline \hline
   $\kappa_c$      & 0.1572(1) & 0.1570(3)  &  0.1569(1) & \\
   $m_{\rho}a$     & 0.343(10) & 0.348(15)  &  0.348(27) & $-$ \\
   $a^{-1}$ (GeV)  & 2.24(7)   & 2.21(10)   &  2.21(17)  & \\
  \hline
   $f_{\rho}/Z_V $ &  0.48(3)  & 0.520(6)   & $-$ & $f_{\rho}=$0.198(5)  \\ 
  \hline
   $ m_{\pi'} / m_{\rho}$ & $-$   &  1.88(8)  & 1.74(8)  & 1.69(13) \\
   $m_{\rho'} / m_{\rho}$ & $-$   &   2.44(11) & 2.25(10) & 
                                   1.90(3)\  {\rm or}\  2.21(3) \\
 \hline
 \end{tabular}
 \end{center}
 \caption{A comparison of the MEM analysis and the asymptotic analysis of the
  same lattice data obtained on the $20^3 \times 24$ lattice with
 $\beta_{lat}=6.0$.  The experimental value of $f_{\rho}$  obtained from
 the decay $\rho \rightarrow e^+e^-$ 
  and the masses of $\rho'$ and $\pi'$ are also shown 
 \cite{PDG}. }
 \label{mass.sum2}
\end{table}

\begin{figure}[hbt]
\vspace*{0.2cm}
\hspace*{-0.5cm}
\epsfxsize=12cm
\centerline{\epsfbox{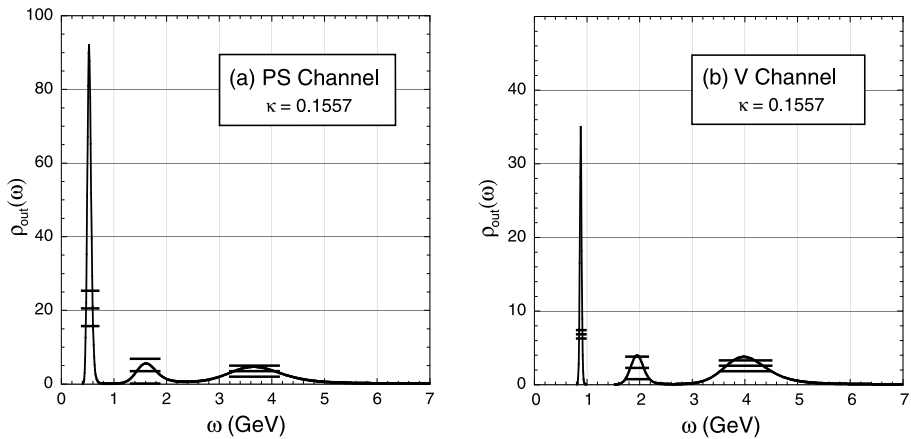}}

\vskip2mm
\caption{$\rho_{out}(\omega)$ in the V and PS channels 
 for $\kappa=$ 0.1557 with error attached. $\omega_{max} = 7.1 $ GeV.}
\label{fig10}
\end{figure}

In Fig.\ref{fig10}, we show
the MEM images in the V and PS channels for $\kappa= 0.1557$
with errors obtained in the procedure discussed in Step 3 in Section 3.3. 
 The meanings of the error bars are the same as those in
 Fig.\ref{fig6}.
 Namely, the horizontal position and length of the 
 bars indicate the frequency region over which
 the SPF is averaged, while the vertical height of 
 the bars denotes the uncertainty in the averaged value
 of the SPF in the interval.

The small error for the lowest peak in Fig.\ref{fig10}
supports our identification of the peak with $\rho$. 
Although the existence of the non-vanishing
spectral strength of the 2nd peak and 3rd bump
is statistically significant, their spectral ``shape''
is either marginal or insignificant.
Lattice data with better quality are called for
to obtain better SPFs.
\footnote{It is in order here to make some comments on the difference
   in the figures shown here and those in our previous publications
   \cite{nah99}.  The lattice data used for the MEM analyses
   are exactly the same in both cases. In \cite{nah99},
   $a^{-1}=2.33 $ GeV (which is obtained on $24^3 \times 54$ lattice
    in the first reference
   of \cite{kc}) was used to set the scale, while in this article we use
   $a^{-1} = 2.21$ GeV obtained from the $\rho$-meson mass 
   in our MEM analysis of the data on the $20^3 \times 24$ lattice.
    This leads to a simple rescaling of
   the factor 0.95 for dimensionful quantities such as 
   $\omega_{max}$ and $\Delta \omega$ from those in \cite{nah99}.
   Also, we mentioned in \cite{nah99} that the averaged heights of the 
    high energy continuum of SPF in the V and PS channels are consistent
    with the perturbative prediction in Table \ref{cont.par}.
    This statement is misleading, since  $Z_j^{PB}$ (which are shown in the
    table) are different from  $Z_j^{DB}$ to be used in our analysis.  }

\begin{figure}[htb]
\vspace*{0.5cm}
\hspace*{-0.5cm}
\epsfxsize=7cm
\centerline{\epsfbox{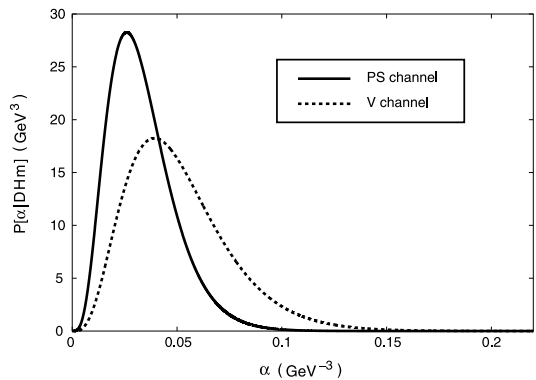}}

\vskip1mm
\caption{$P[\alpha|DHm]$ for the PS and V channels.}
\label{fig11}
\vspace*{0.5cm}
\end{figure}

In Fig.\ref{fig11}, $P[\alpha|DHm]$ 
   (with the  approximation discussed in Step 2 in Section 3.3) 
is shown for the PS and V channels in the case of $\kappa= 0.1557$.
 We have found that the SPFs obtained after averaging over $\alpha$ and
 those at $\alpha = \hat{\alpha}$ have negligible difference
 although the distribution of $\alpha$ spreads over the 
 range $0 \rightarrow 1$ in Fig.\ref{fig11}.  

The sensitivity of the 
results under the variation of $\omega_{max}$ is shown 
in Fig.\ref{fig12}, where $\omega_{max} = 8.5$ GeV
is adopted.  Comparison of Fig.\ref{fig12} 
 and  Fig.\ref{fig9} shows no appreciable change of SPFs under the
 variation of $\omega_{max}$ as long as 
 $\omega_{max} \ge \pi /a$.

We have also checked that 
the result is not sensitive, within the statistical
significance of the image, to the variation
of $m_0$ by factor 5 as shown in Fig.\ref{fig13}.
 The default model dependence is relatively weak compared to 
 the case of the mock data shown in Fig.\ref{fig6} partly
 because the off-diagonal components of the covariance matrix
 for the lattice data are not negligible and stabilize the
 final images.

\newpage
\begin{figure}[htb]
\hspace*{-0.5cm}
\epsfxsize=12cm
\centerline{\epsfbox{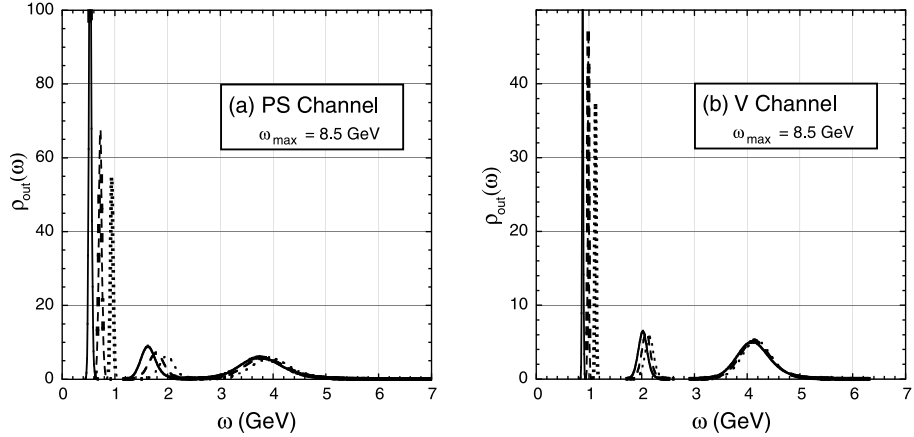}}

\vskip1mm
\caption{SPFs in the V and PS channels for $\omega_{max}=8.5$ GeV.
The solid, dashed and dash-dotted
lines are for $\kappa=$ 0.1557, 0.1545 and 0.153, respectively.
}
\label{fig12}
\vspace*{-0.3cm}
\end{figure}

\begin{figure}[htb]
\vspace*{0.5cm}
\hspace*{-0.5cm}
\epsfxsize=6cm
\centerline{\epsfbox{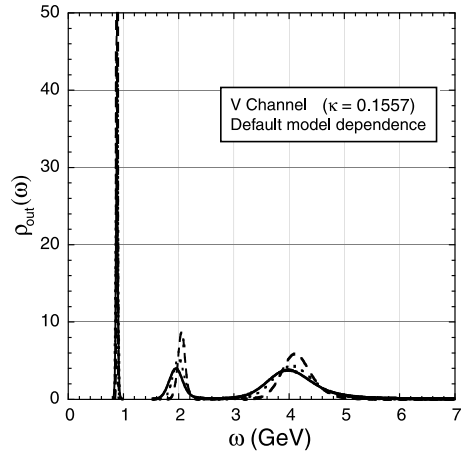}}

\vskip2mm
\caption{The default model dependence of 
 the output image in the vector channel for  $\omega_{max}=7.1$ GeV.
 The solid line corresponds to  $m_0 = 0.6$ (the same value
 with that taken in Fig.\ref{fig9} and Fig.\ref{fig10}, 
 while the long (short) dashed line corresponds to
 $5 m_0$ ($m_0 /5$).}
\label{fig13}
\vspace*{0.1cm}
\end{figure}

\subsubsection{Scalar and axial-vector channels}

In Fig.\ref{fig14}, the spectral functions for
 the S and AV channels are shown.
  In those channels, we have used  $\tau_{max}/a =12~(10)$ 
 for the S (AV) channel. Smaller $\tau_{max}$ is chosen 
 in the AV channel 
  to avoid the pion contamination 
 that contributes to the correlator at large $\tau$ 
 as discussed in Section 5.3.
 Although the peak + continuum structure is seen in
  Fig.\ref{fig14},
 SPF does not change in a regular way under the
 variation of the quark mass, which prevents us
 from making chiral extrapolation of the peaks.
 More statistics and also better technique are needed to obtain
 SPFs with good quality in these channels.
 
\begin{figure}[htb]
\vspace*{0.5cm}
\hspace*{-0.5cm}
\epsfxsize=12cm
\centerline{\epsfbox{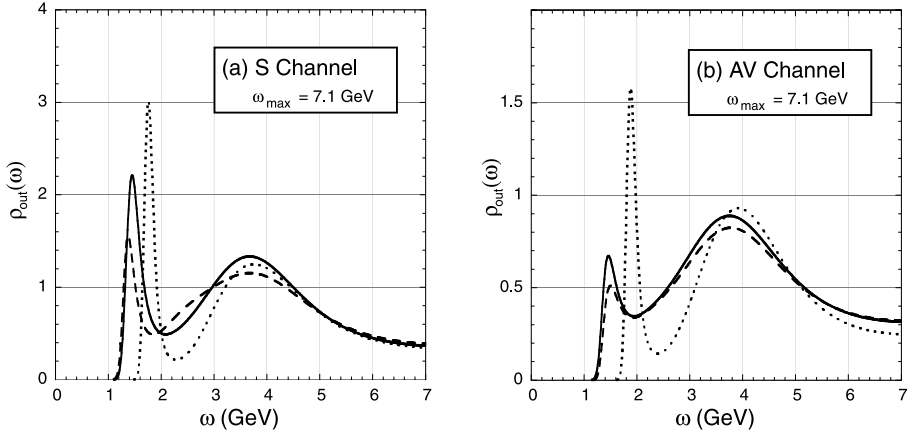}}

\vskip2mm
\caption{SPF for S and AV channels.
 The solid, dashed and dash-dotted
lines are for $\kappa=$ 0.1557, 0.1545 and 0.153, respectively.
	}
\label{fig14}
\vspace*{0.3cm}
\end{figure}

\subsubsection{Lattice versus continuum kernel}

In the MEM analysis presented so far,
we have relied on the spectral representation
 derived in the continuum limit,
\beq
\label{dis-cont}
D(\tau) = \int_0^{\infty} \ K(\tau, \omega)
\ A(\omega) \  d\omega .
\eeq
Here the kernel $K$ at $T=0$ is related to the 
free boson propagator with a mass $\omega$:
\beq
 \label{k-cont}
K(\tau, \omega)  = e^{-\omega \tau} 
 =  2 \omega \int_{-\infty}^{+\infty}
 {d\nu  \over 2 \pi} {e^{i\nu \tau} \over \omega^2 + \nu^2}.
\eeq

Now, to study the systematic error from the finite lattice
 spacing in extracting SPF, 
 one may artificially define a ``lattice spectral representation'' as
\beq
\label{dis-lat}
D(\tau) = \int_0^{\infty} \ K^{lat}(\tau, \omega) 
 \ A^{lat}(\omega) \ d\omega ,
\eeq
where $K^{lat}$ is a ``lattice kernel'' at $T=0$ defined through the 
 lattice boson propagator with a mass $\omega$:
\beq
\label{k-lat}
K^{lat}(\tau, \omega; a) = 2 \omega \int_{-\pi /a}^{+\pi /a}
 {d\nu  \over 2 \pi} {e^{i\nu \tau} \over 
  \omega^2 + ({2\over a} \sin {\nu a \over 2})^2}.
\eeq

 $K^{lat}$ and $A^{lat} $ approach $K$ and $A$, respectively,
 in the continuum limit. Therefore, by taking the same
 lattice data $D(\tau)$ on the l.h.s.
 of (\ref{dis-cont}) and (\ref{dis-lat}) and by comparing
 the resultant  $A(\omega)$ and $A^{lat}(\omega)$ in the MEM analysis,
 one can estimate a part of the systematic error from the finiteness
 of the lattice spacing.

 The difference of
  $K^{lat}$ and $K$ becomes substantial for large $\tau \omega$,
 which is  shown in Fig.\ref{fig15}. Therefore, 
 the finite $a$ error appears typically at large $\omega$.
 The spectral functions in the V and PS channels obtained with the lattice 
 kernel are shown in Fig.\ref{fig16}.
 In Table \ref{mass.sum2}, the masses extrapolated to the chiral limit
 are shown in the column ``MEM lattice kernel''.
  ($f_{\rho}$ is not shown in the table, since it is difficult to
 isolate the $\rho$-pole unambiguously from Fig.\ref{fig16}.)
 The results  are  consistent with those 
 of ``MEM continuum kernel" within error bars.
 In Fig.\ref{fig17}, the SPFs with error attached are shown
 for the PS and V channels. Taking into account those errors,
 it is hard to distinguish the images obtained by
 $K$ and $K^{lat}$ with the present lattice data, although the
 SPFs in Fig.\ref{fig17} are flat compared with those
 in Fig.\ref{fig10}.
     
\begin{figure}[htb]
\vspace*{0.5cm}
\hspace*{-0.5cm}
\epsfxsize=6cm
\centerline{\epsfbox{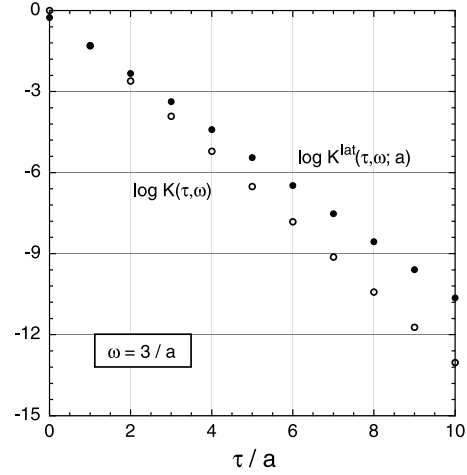}}

\vskip2mm
\caption{Comparison of the lattice kernel $K^{lat}$ and
  the continuum kernel $K$ 
 near the highest frequency on the lattice $\omega = 3/a$.}
\label{fig15}
\vspace*{0.1cm}
\end{figure}

\begin{figure}[htb]
\vspace*{0.5cm}
\hspace*{-0.5cm}
\epsfxsize=12cm
\centerline{\epsfbox{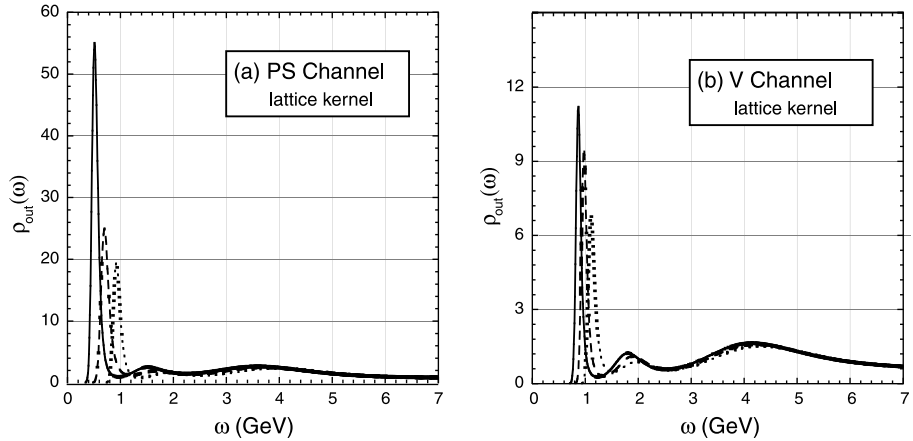}}
\vskip2mm
\caption{SPFs for the V and PS channels obtained with the lattice kernel.
The solid, dashed and dash-dotted
lines are for $\kappa=$ 0.1557, 0.1545 and 0.153, respectively.
}
\label{fig16}
\vspace*{0.3cm}
\end{figure}

\newpage

\begin{figure}[htb]
\hspace*{-0.5cm}
\epsfxsize=12cm
\centerline{\epsfbox{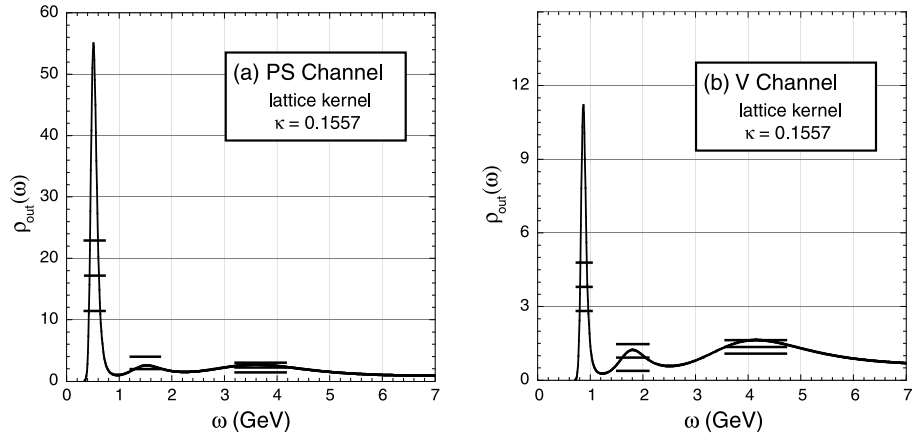}}
\vskip2mm
\caption{SPFs for the V and PS channels
obtained with the lattice kernel with error attached
for $\kappa=0.1557$. }
\label{fig17}
\end{figure}

\newpage

\setcounter{equation}{0}
\setcounter{subsection}{0}

\section{Summary and concluding remarks}

In this article, we have examined the spectral function (SPF) in QCD and
its  derivation from the lattice QCD data.
The maximum entropy method (MEM), 
which allows us to study SPFs without
making a priori assumptions on the spectral shape, turns out to be
quite successful and promising even for 
limited number of  lattice QCD data with noise.
In particular, the uniqueness of the solution 
and the quantitative error analysis make MEM
superior to any other approaches adopted previously
for studying SPFs on the lattice. Also,  
the singular value decomposition (SVD) 
of the kernel of the spectral sum rule leads to a 
very efficient algorithm for obtaining SPFs numerically.
    
 By analyzing  mock data, 
 we have tested  that the MEM image approaches the exact one
 as the number of temporal data points is increased and the statistical error
 of the data is reduced.
 Even with the lattice QCD data at
 $T=0$ obtained on the $20^3 \times 24$ lattice
 with the lattice spacing $a \simeq 0.09$ fm, MEM was able to 
 produce resonance peaks with correct masses
 and the continuum structure.
 The statistical significance of the obtained spectral functions
  has been also analyzed. 
 Better data with smaller $a$ and larger lattice volume will
 be helpful for obtaining  SPFs with smaller errors.  

\vspace{0.5cm}

 MEM introduces 
 a new way of extracting physical
  information as much as possible   from the lattice QCD
  data.  Before ending this article, we list 
 open problems for future studies. Some of them are 
 straightforward and some of them require more work.

\vspace{0.5cm}

\noindent
Technical issues:

\begin{itemize}

  \item[(1)] The number of temporal points $N$ and the lattice spacing $a$
        are  the crucial quantities
        for the MEM analysis to be successful.  In this article, we have
        fixed $N= 12$ and $a=0.09$ fm  for extracting SPF 
          from the lattice  data. However, it is absolutely necessary
       to study  $N$ and $a$ dependence of the resultant SPFs.  
        For this purpose, we have collected 160 gauge configurations
        on CP-PACS at Univ. of Tsukuba with $40^3 \times 30$ lattice
        and $\beta_{lat} = 6.47$ ($a\sim 0.05$ fm) with the periodic
        boundary condition.  The detailed MEM analysis of the data
         will be reported elsewhere \cite{ahn_finer}.   
 
\item[(2)]  We have made chiral extrapolation only for the 
        peaks of SPFs but not for the whole spectral structure in this article:
        In fact we have found that neither the
        direct extrapolation of the MEM image $A_{out}(\omega)$ 
        nor the extrapolation of $D(\tau)$ and $C$
        works in a straightforward manner.
        This is an open problem for future study. 
 
\item[(3)] Although we have shown that one can select 
        an optimal default model
        by varying $m(\omega)$ and estimating the errors of SPFs,
        the variation was within an assumed functional form
        such as $m(\omega) = m_0 \omega^n$ motivated by the
        asymptotic behavior of the spectral functions in QCD. 
        How to optimize the default model
        given data in a systematic way should be   studied further.
        (For a possible procedure by estimating $P[m|DH]$, see 
        ref.\cite{physrep}.)

\end{itemize}

\newpage

\noindent
Physics issues:

\begin{itemize}
 
  \item[(4)] Further studies on the scalar and axial-vector
         channels are necessary to explore the
         applicability of MEM for not-so-clean lattice-data. 
 
  \item[(5)] MEM analysis   for baryons is quite useful in
  extracting information on excited baryons and their
  chiral structure \cite{ahn_baryon}.         

  \item[(6)] Applications to heavy resonances such as the glueballs,
         and charmed/bottomed hadrons 
         will be an ideal place for MEM, since one
         can extract ground state peaks with limited number of data
         points obtained at relatively short distances.
         For the study of charmonium
         on $40^3 \times 30$ lattice
	 with $\beta_{lat}=6.47$, see \cite{ahn_finer}.
 
  \item[(7)]  When one studies  a two-point correlation function   of 
        operators ${\cal O}_1$ and ${\cal O}_2$ with
         ${\cal O}_1\neq {\cal O}_2$, 
        one encounters  SPF which is not necessarily
	positive semi-definite.
        Typical examples are the meson and baryon mixings such
         as   $\rho$-$\omega$, $\pi^0$-$\eta$,
         $\eta$-$\eta'$,  glueball-$q\bar{q}$ and 
         $\Lambda$-$\Sigma^0$  (see e.g. \cite{hmhk}).
         The generalization of the Shannon-Jaynes entropy to non
          positive-definite images is possible \cite{ahn_non-pos},
	 and it is an interesting
         future problem to study hadron mixings from the generalized MEM.

  \item[(8)] Full QCD simulations combined with MEM may open 
      up a possibility of
         first principle determination of resonance widths such 
         as $\rho(770)\rightarrow 2\pi$, $\sigma(400-1200) \rightarrow 2 \pi$
          and $a_1 (1260) \rightarrow 3 \pi$. Also it serves for
         unraveling the structure of the mysterious scalar meson ``$\sigma$''
        \cite{sigma-meson}.
 
  \item[(9)] The long-standing problem of in-medium spectral functions of
        vector mesons
         ($\rho$, $\omega$, $\phi$, $J/\psi$, $\Upsilon$, $\cdots$, etc ) 
        and scalar/pseudo-scalar mesons ($\sigma$, $\pi$, $\cdots$, etc.)
        can be studied using MEM combined with finite $T$ lattice simulations.
        The in-medium behavior of
        the light vector mesons \cite{ak93,rw99,HL,Lee-rev,pisarski,br} and
        scalar mesons \cite{hk85,ch98} is intimately
        related to the chiral
        restoration in hot/dense matter, while that of the 
        heavy vector mesons 
        is related to deconfinement \cite{JP-review,HM,MS}.
        An anisotropic lattice is necessary for this purpose
        to have enough
        data points in the temporal direction at 
        finite $T$ \cite{ahn_an-iso}. (See also a recent attempt
        on the basis of NRQCD simulation \cite{ukqcd_ohio}.)
        Also, it is interesting to study
        SPFs with finite three-momentum ${\bf k}$,
        since $\omega$ and ${\bf k}$ enter
        in the in-medium SPFs as independent variables.

  \item[(10)] Possible existence of non-perturbative  
        collective modes above the critical temperature of 
        the QCD phase transition speculated in  \cite{hk85,DeTar,thoma}  
        may also be studied efficiently by using MEM, 
        since the method does not
        require any specific ans\"{a}tze for the spectral shape.
        Also, correlations  in the diquark channels in the vacuum
        and in medium are 
        interesting to be explored in relation to the 
        $q$-$q$ correlation
        in baryon spectroscopy and to color superconductivity
        at high baryon density \cite{wk}. 
         
 \end{itemize}

\newpage

\noindent
\begin{center}
{\bf Acknowledgements}
\end{center}

We are grateful to MILC collaboration for providing us their open codes
for lattice QCD simulations, which has enabled this research.
 Most of our simulations presented in this article
  were carried out on a Hitachi 
SR2201 parallel computer at Japan Atomic Energy Research Institute.
 We thank S. Chiba for giving us an opportunity to use the above computer.
 We also thank S. Aoki, K. Kanaya, A. Ukawa and T. Yoshie for
 their interests in this work and encouragements, and thank K. Sasaki 
  for reading the manuscript carefully.
  T. H. thanks H. Feldmeier
for the discussion  on
the uniqueness of the solution of MEM.
M. A. (T. H.) was partly supported by Grant-in-Aid for Scientific
Research No. 10740112 and No. 11640271 (No. 98360 and No.12640263)
of the Japanese Ministry of Education,
 Science and Culture.

\newpage

\appendix 


\section{Monkey argument for entropy and  prior probability}

\renewcommand{\theequation}{A.\arabic{equation}}

 Throughout Appendix A and B, we shall use the common
 notation $f(x)$ for the 
 image instead of $A(\omega)$ used in the text.
  
 What we  need in the maximum entropy method
 is the prior probability $P(f \in V)$, namely,
 the probability that the image $f(x)$ is in a certain
 domain $V$. It can be generally written as
\beq
\label{pr2}
P(f \in V)  =  {1 \over Z_S(\alpha)}
 \int_V [df] \  \Phi(\alpha S(f)),
\eeq
where $\alpha$ is an arbitrary constant defined for later use,
and $Z_S(\alpha)$ is a normalization factor. We
assume that $\Phi$ is a monotonic function of the would-be entropy 
$S(f)$, so that the most probable image $f$ is obtained as a
stationary point of $S(f)$.
                 
 The so-called ``monkey argument'', which is based on the
 law of large numbers, can determine the
 explicit form of $\Phi$ and   $S(f)$ 
 (see, e.g., \cite{wu,skill1,gd78,jaynes86}). Here
 we shall recapitulate  the essential part of this derivation.

 Let us first discretize the basis space $x$ into
 $N$ cells. Correspondingly, $f(x)$ is discretized as $f_i$
 $(1 \leq i \leq N)$.
 Suppose a monkey throws $M$ balls.
 $M$ is assumed to be large.
 Define $n_i$ as the actual number of balls which the $i$-th cell
 received. Also, the $i$-th cell has a probability $p_i$ 
 to receive a ball.  Then $\lambda_i$ (the
 expectation value of the number of the balls in the $i$-th cell) reads
 $ \lambda_i = M p_i $ with $ \sum_{i=1}^N \lambda_i = M$.

The probability that the $i$-th cell receives $n_i$ balls
 is given by the binomial distribution. Its
 large $M$ limit with $\lambda_i$ fixed is the
 Poisson distribution $P_{\lambda_i}(n_i)$:
\beq
B_{M,\, p_i}(n_i) & = & \ \frac{M !}{n_i ! (M-n_i)!}\
   p_i^{n_i}\  (1-p_i)^{M-n_i}
\rightarrow  P_{\lambda_i}(n_i)   \ \ \ (M \rightarrow \infty ).
\eeq
     
 Therefore,  the probability that a certain image
 $\vec{n} =( n_1, n_2, \cdot \cdot \cdot , n_N )$ is realized
 reads
\beq
\label{pr3}
 P_{\vec{\lambda}}(\vec{n} )
  = \prod_{i=1}^N P_{\lambda_i}(n_i )
  = \prod_{i=1}^N \ { \lambda_i^{n_i} \
e^{-\lambda_i} \over n_i ! },
\eeq
where the normalization is given by
$\sum_{n_i=0}^{\infty} P_{\lambda_i}(n_i) =1$ 
$(i=1,2,\cdots,N)$.

 Since $n_i$ may be large as $M$ is large,
 we  introduce  a small ``quantum'' $q$ and define
 the finite image $f_i$ and the default model $m_i$ as
\beq
\label{pr4}
f_i = q n_i, \quad m_i = q \lambda_i.
\eeq

In terms of (\ref{pr4}), $P(f \in V)$ is written as
\beq
\label{pr50}
P(f \in V)
 =  \sum_{\vec{n} \in V }
 P_{\vec{\lambda}} (\vec{n})
\label{pr51}
\simeq
 \int_V {\prod_{i=1}^N df_i \over q^N } \
 \prod_{i=1}^N { \lambda_i^{n_i}\  e^{-\lambda_i} \over n_i ! }
\simeq 
 \int_V \prod_{i=1}^N {df_i \over \sqrt{ f_i }}
\   {e^{S(f)/q} \over (2 \pi q)^{N/2}} ,
\eeq
where small $q$ is assumed for converting the sum to
 the integral, and the Stirling's formula,
 $n!  \simeq \sqrt{2\pi n} e^{n \log n -n} $,
 is used to obtain the last expression.

  $S(f)$ is nothing but the Shannon-Jaynes entropy,    
\beq
 S(f) = \sum_{i=1}^N \left[ f_i - m_i
 - f_i \log \left( {f_i \over m_i} \right) \right] .
\eeq
 
Comparing (\ref{pr51}) with (\ref{pr2}),
 one finds
\beq
\label{pr7}
q = \alpha^{-1}, \ \ \ [df] = \prod_{i=1}^N
\frac{df_i}{\sqrt{f_i}}, \ \ \
 Z_S(\alpha) = \left ( \frac{2 \pi}{\alpha }\right )^{N/2} .
\eeq

 Note also that the measure $M(f)$ defined by
\begin{equation}
M(f)= \prod_{i=1}^N {f_i}^{-1/2}
\end{equation}
is rewritten as
\beq
\label{pr8}
M(f) = \sqrt{ \det g } \ \ \ {\rm with} \ \ \
g_{ij} = (1 /f_i ) \delta_{ij}
= -{ \partial^2 S(f) \over \partial f_i \partial f_j } .
\eeq
 The metric tensor $- g_{ij}$ for the functional
 integral over $f$ is thus related to the curvature matrix of $S(f)$.

\section{Axiomatic construction of entropy}

\renewcommand{\theequation}{B.\arabic{equation}}

 The axiomatic construction of the Shannon-Jaynes entropy $S$
  given below is a modified
 version of that given in  \cite{skill2}.
 We have changed  the statement of each axiom so that
 it becomes easier to understand the idea behind.

    For positive semi-definite distribution $f(x)$, we want to
     assign a real
number $S(f)$ (the Shannon-Jaynes entropy) as
\beq
\label{image}
   f \mbox{ is a more plausible image than } g 
  \ \ \leftrightarrow 
  \ \ S(f) > S(g) .
\eeq
 If there exists an external
 constraint on $f(x)$ such as $C(f(x))=0$, 
 the most plausible image is
 given by the following condition
\beq
\delta_f \left[ S(f) - \lambda C(f) \right]  =0,
\eeq
with $\lambda$ being a Lagrange multiplier.
The explicit form of $S$ is uniquely fixed by the following axioms.

\vspace{0.5cm}

\noindent
{\bf Axiom I: Locality}

\noindent $S(f)$ is a local functional of
 $f(x)$ without derivatives.
 Namely, there is no correlation between the images at different $x$.

 This leads to a form
\beq
\label{ax1}
S(f) = \int dx \ m(x) \ \theta(f(x),x).
\eeq
Here $m(x)$ is a positive definite function which
 defines the integration measure.
 $\theta$ is an arbitrary local function of $f(x)$ and $x$
 without derivatives acting on $f$.

\vspace{0.5cm}

\noindent
{\bf Axiom II: Coordinate Invariance}

\noindent $f(x)$ and $m(x)$ transform as
 scalar densities under the
 coordinate transformation $x'=x'(x)$, namely,
 $f(x) dx = f'(x') dx'$ and $m(x) dx = m'(x') dx'$. 
 Also, $S$ is a scalar.  
 
 This axiom allows only  two invariants  for constructing $S$ in
 (\ref{ax1}):  $ m(x)dx = m'(x')dx'$ and $f(x)/m(x) = f'(x') /m'(x')$.
 Therefore, 
 \beq
\label{ax2}
S(f) = \int dx \ m(x) \ \phi(f(x)/m(x)) .
\eeq

\vspace{0.5cm}

\noindent
{\bf Axiom III: System Independence}

\noindent If  $x$ and $y$ are two independent variables,
 the image $F(x,y)$  is written as a {\em product} form
  $F(x,y) = f(x) g(y)$  together
   with the integration measure $m(x,y) = m_f(x) m_g(y)$.
 Furthermore,   
 the first variation of $S(F)$ with respect to
  $F(x,y)$  leads to an {\em additive} form with some
  functions $\alpha(x)$ and $\beta(y)$;
\beq
\label{add-rule}
 { \delta S(F) \over \delta F(x,y) } = \alpha(x) + \beta(y).
 \eeq

From this axiom, 
images $f(x)$ and $g(y)$ are determined independently 
 from the variational equation
  ${ \delta S(F) / \delta F(x,y) }=0$. 
 First of all, (\ref{ax2}) reads
\beq
\label{p3}
 S(F)  =  \int dx \int dy \ m(x,y) \ \phi (F(x,y)/m(x,y)) .
\eeq
Acting  the derivative $\partial^2 /\partial x \partial y$
on eq.(\ref{add-rule}) with  (\ref{p3})
 and using
 $Z \equiv F(x,y)/m(x,y) = (f(x)/m_f(x))(g(y)/m_g(y))$, one obtains
\beq
Z {d^2 \sigma(Z) \over dZ^2} + {d  \sigma(Z) \over dZ }=0,
\eeq
where $\sigma(Z) = d\phi (Z) /dZ$.
The solution of this differential equation is
$\sigma(Z)=c_1 \log Z - c_0$, which
leads, up to an irrelevant constant, to
\beq
\phi(Z) = c_1 Z \log Z - (c_0+c_1) Z.
\eeq
Thus one arrives at
\beq
\label{ax3}
S(f) = \int dx \ m(x) \phi(f/m) = \int dx \ f(x) 
 \left[ c_1 \log \left( {f(x) \over m(x) } \right) - (c_0 + c_1) \right] ,
\eeq
where we have dropped the suffix $f$ in $m_f(x)$ for simplicity.
 The curvature of $S$ is dictated by $c_1$,
 since $(\delta /\delta f)^2  S(f) = c_1 / f $ and  $f \ge 0$.
  We choose
 $c_1=-1$ for having  $S$ bounded from above 
  and for overall normalization.

\vspace{0.5cm}

\noindent
{\bf Axiom IV: Scaling}

\noindent If there is no external constraint on $f(x)$, 
 the initial measure is recovered after the variation, i.e., $f(x) = m(x)$.
 
 This axiom leads to $c_0 = 0$ in (\ref{ax3}), since the
 unconstrained solution of  $\delta  S(f) /\delta f =0$ is
 $f(x) = m(x) e^{c_0/c_1}$.
  Also, it is convenient to add a constant to $S(f)$ in
 (\ref{ax3}) to make $S(f=m)=0$.
Thus we arrive at
\beq
\label{ax4}
S(f) = \int dx \left[ f(x) - m(x) - f(x) \log \left( {f(x)
 \over m(x)} \right)  \right] ,
\eeq
which  is the desired  expression of the Shannon-Jaynes entropy.

\section{The singular value decomposition}

\renewcommand{\theequation}{C.\arabic{equation}}

Here we give a proof of the singular value decomposition (SVD) of a general 
$m \times n$ matrix $P$ following \cite{chaterin}.

The singular values of a matrix $P$ are defined as 
 the square root of the  eigenvalues of $P^{\dagger}P$.
 By definition,  $P^{\dagger}P$ is 
 an $n \times n$ Hermitian matrix and has real and non-negative
 eigenvalues.
 We also define the norm of the vector $x \in {\bf C}^n$ and
 the spectral norm of the $m \times n$ matrix $P$, respectively, as
\beq
\parallel x \parallel_2 & = &  \left[ \sum_{i=1}^n |x_i |^2 \right]^{1/2}, \\
\parallel P \parallel_2 & = & 
\left[ {\rm maximum\  eigenvalue\ of\ } P^{\dagger}P \right]^{1/2} \\
& = & \left[ {\rm max}(x^{\dagger} P^{\dagger}P x) \right]^{1/2} \ \ 
(x \in {\bf C}^n, \parallel x \parallel_2 =1).
\eeq

\noindent {\bf SVD Theorem}:\\
Let $P$ be an $m \times n$ matrix ($m \ge n$),
 $U$ be an $m \times m$ unitary matrix, and 
 $V$ be an $n \times n$ unitary matrix.
 Then, $P$ can be decomposed as
\beq
P = U \Xi V^{\dagger},
\eeq
where $\Xi$ is an $m \times n$ diagonal matrix
 with the diagonal elements being the singular values of 
 $P$, namely, $\Xi = {\rm diag}(\xi_1, \xi_2, \cdots, \xi_n)$
 with $\xi_1 \ge \xi_2 \ge \cdots \ge \xi_n \ge 0$.

\noindent {\bf Proof}:\\
Define the maximum singular value of $P$ as $\xi_1$.
 Then, there exists a vector $x_1$ which satisfies
 the relation: $\xi_1^2 = (x_1^{\dagger} P^{\dagger}P x_1)$.
 Also there exists  a vector 
$ y_1 \in {\bf C}^m ( \parallel y \parallel_2 =1)$ with the property,
 $P x_1 = \xi_1 y_1$.

For $x_1$ and $y_1$ obtained above, one may introduce non-square matrices
 $U_2$ and $V_2$ such that 
 $U_1$ and $V_1$ given below become an $m \times m$ unitary matrix
  and an $n \times n$ unitary matrix, respectively:
\beq
U_1 = ( y_1, U_2 ), \ \ \ V_1 = (x_1, V_2) .
\eeq

Using $U_1$ and $V_1$ defined above, $P$ is transformed into $P_1$
as 
\beq
P_1 \equiv U_1^{\dagger} P V_1 = 
\left( \begin{array}{c} y_1^{\dagger} \\ U_2^{\dagger} \end{array} \right)
 P (x_1, V_2) 
= \left( \begin{array}{cc} \xi_1 & y_1^{\dagger}PV_2 \\
                            0    & U_2^{\dagger} P V_2 \\ \end{array}
  \right) 
\equiv  
\left( \begin{array}{cc} \xi_1 & z_1^{\dagger} \\
                            0  & Q_2 \\ \end{array} 
\right ) ,
\eeq
where $z_1 \in {\bf C}^{n-1}$ and $Q_2$ being an $ (m-1) \times (n-1)$
 matrix.

Now, we show that $z_1$ is actually a null vector:
\beq
\xi_1^2 & = & \parallel P \parallel_2^2 
          = \parallel P_1 \parallel_2^2
          = {\rm max} (x^{\dagger} P_1^{\dagger} P_1 x) 
                      \nonumber \\   
        & \ge & 
          { 1 \over \xi_1^2+ \parallel z_1 \parallel_2^2}  \ 
                (\xi_1, z_1^{\dagger}) P_1^{\dagger} P_1 
                 \left( \begin{array}{c} \xi_1 \\  z_1 \\ \end{array}
                 \right)            \nonumber     \\
        & = & { 1 \over \xi_1^2+\parallel z_1 \parallel_2^2}
               \left[ (\xi_1^2+\parallel z_1 \parallel_2^2)^2 + 
                 \parallel Q_2 z_1 \parallel_2^2 \right] \nonumber \\
        & \ge & \xi_1^2+\parallel z_1 \parallel_2^2.
\eeq
Since $\xi_1$ is the maximum singular value of $P$ and $P_1$,
 $\xi_2$ defined as the maximum SV of $Q_2$ satisfies
   $\xi_1 \ge \xi_2 $.
   Applying the same procedure to $Q_2$, $Q_3$, $\cdots$,  one
   finds
\beq
P_n =  U^{\dagger} P V = 
\left( \begin{array}{ccc} 
       \xi_1 &        &       \\
             & \ddots &       \\
             &        & \xi_n \\
         0   & \cdots &   0   \\      \end{array}   
\right) \equiv \Xi .
\eeq
Thus one arrives at the singular value decomposition $P = U \Xi V^{\dagger}$.
  (QED)

One may neglect the irrelevant components of $U$ and $\Xi$ so that
they are an $m \times n$ matrix and an $n \times n$ matrix, respectively.
In this case, $U$ satisfies the condition $U^{\dagger}U =1$, while
$V^{\dagger}V = V V^{\dagger}=1$. This form of SVD is used in the text.


\end{document}